\documentclass[article]{aa} 
\usepackage{multirow}
\usepackage{multicol}

\usepackage{bm}
\usepackage{imakeidx} 
\usepackage[T1]{fontenc}
\usepackage{siunitx} 
\usepackage{graphicx} 
\usepackage[table,xcdraw]{xcolor} 
\usepackage{caption}
\usepackage{subcaption}
\usepackage{amsmath} 
\usepackage{geometry}
\usepackage{float} 
\usepackage{here} 
\usepackage{xcolor} 
\usepackage{inputenc}
\usepackage[colorlinks=true, pdfstartview=FitV, linkcolor=blue, citecolor=blue, urlcolor=blue, breaklinks=true]{hyperref}

\usepackage{hyphenat}

\def\Teff{T_{\rm eff}}

\def\deg{\ensuremath{^\circ}}

\newcommand{\Gaia}{{Gaia}}
\newcommand{\gspspec}{{GSP-Spec}}

\newcommand{\mh}{[M/H]}
\newcommand{\feh}{[Fe/H]}

\newcommand{\alphafe}{[$\alpha$/Fe]}

\newcommand{\cafe}{[Ca/Fe]}
\newcommand{\cah}{[Ca/H]}

\newcommand{\mgfe}{[Mg/Fe]}

\newcommand{\cefe}{[Ce/Fe]}

\newcommand{\xfe}{[X/Fe]}

\usepackage{color}

\usepackage{cleveref}
\usepackage{tabularx,booktabs}
\usepackage{fancyvrb}

\begin{document}


\title{\textbf{Constraints on the history of Galactic spiral arms revealed by $\Gaia$ $\gspspec$ $\alpha$-elements}}

\author{M. Barbillon \inst{1}, A. Recio-Blanco \inst{1}, E. Poggio \inst{1,2}, P. A. Palicio\inst{1}, E. Spitoni\inst{3,1}, P. de Laverny \inst{1}, G. Cescutti \inst{4,3,5}} 
\institute{\inst{1}\textit{Université Côte d’Azur, Observatoire de la Côte d’Azur, CNRS, Laboratoire Lagrange, France}\\
\inst{2}\textit{Osservatorio Astrofisico di Torino, Istituto Nazionale di Astrofisica (INAF), 10025 Pino Torinese, Italy}\\
\inst{3}\textit{I.N.A.F. Osservatorio Astronomico di Trieste, via G.B. Tiepolo 11, 34131, Trieste, Italy}\\
\inst{4}\textit{Dipartimento di Fisica, Sezione di Astronomia, Università di Trieste, Via G. B. Tiepolo 11, 34143 Trieste, Italy}\\
\inst{5}\textit{INFN, Sezione di Trieste, Via A. Valerio 2, I-34127 Trieste, Italy}
}

\date{Received 24 may 2024; Accepted 8 november 2024}


\abstract{The distribution of chemical elements in the Galactic disc can reveal fundamental clues on the physical processes that led to the current configuration of our Galaxy.}
{We map chemical azimuthal variations in the disc using individual stellar chemical abundances like calcium and magnesium, and discuss their possible connection with the spiral arms and other perturbing mechanisms.} 
{Using Gaia Data Release 3, we examine [Ca/Fe] and [Mg/Fe] fluctuations in a $\sim$4 kpc region around the Sun, focusing on bright giant stars. We implemented a kernel density estimator technique to enhance the chemical inhomogeneities.}
{We observe radial gradients and azimuthal fluctuations in [$\alpha$/Fe] for young ($\lessapprox$ 150 Myr) and old ($\gtrapprox$ 2 Gyr) stars, with amplitudes varying according to the studied element. In young stars, those within spiral arms (e.g., Sagittarius-Carina and Local arms) are generally more metal and calcium-rich ($\sim$0–0.19 dex) but show lower [Ca/Fe] ($\sim$0.06 dex) and [Mg/Fe] ($\sim$0.05 dex) compared to inter-arm regions, suggesting enhanced iron production in spiral arms. These [alpha/Fe] depletions are analysed in light of theoretical scenarios and compared to a 2D chemical evolution model with multiple spiral patterns. For the old sample, [Ca/Fe] maps reveal deficiencies along a segment of the Local arm identified by young stars. We caution that, for this old sample, the quality of the obtained maps might be limited along a specific line-of-sight, due to the Gaia scanning law.}
{This study transitions our understanding of disc chemical evolution from a 1D radial view to a more detailed 2D framework incorporating radial, azimuthal, and small-scale variations. Individual chemical abundances prove essential for tracing spiral arms in disc galaxies. We recommend models and simulations incorporate alpha-abundance trends to better address spiral arm lifetimes.

}


\titlerunning{Constraints on the history of Galactic spiral arms revealed by $\Gaia$ $\gspspec$ $\alpha$-elements}
\authorrunning{Barbillon et al.}

\maketitle
\section{Introduction}
\label{Introduction}


The ability to map the structural, kinematic and chemical properties of stars in our Galaxy has revealed that the classical components of the Milky Way (MW), namely the disc(s) \citep[composed of the thin and thick discs][]{thin_disc_scale}, the bulge and the halo, are interlinked and constitute a system interacting with its environment \citep[e.g.][]{Gaia-enceladus, phase_spiral_DR3}. The accretion of satellite galaxies can leave important signatures in the Galactic disc \citep{Purcell:2011,Laporte:2018,GaravitoCamargo:2021}. Moreover, the large-scale spiral structure of the Galaxy \citep[e.g.][]{MW_structure1,spiral_arm_detection7,MW_structure6,MW_structure4} and the central bar \citep[e.g.][]{MW_structure2,MW_structure7, MW_structure5, DR3_asymmetry_disc} are expected to be crucial drivers of the dynamical evolution of disc stars \citep[e.g.][ and others]{churning_blurring, Minchev:2012, Hunt_2019, radial_migration_3, spiral_arm_radial_action}.


As a consequence of the dynamical mechanisms at work in the Galactic disc, the star kinematic history can
be completely erased. On the other hand, the chemical composition will be preserved through the stellar life. In a galaxy with a radial metallicity gradient \citep{Anders_2017, Katz_2021}, any process of radial migration would give rise to chemical azimuthal variations.

To disentangle the dynamical history of Galactic stellar populations, mapping the spatial distribution of metals in the disc has a crucial importance. In external disc galaxies, both radial and azimuthal chemical gradients have been mapped \citep{Pilyugin:2014,azimuthal_gradient_external_4, Zurita:2021}. In the MW, chemical azimuthal variations have been found using HII regions \citep{metallicity_gradient_3,Wenger:2019}, Cepheids \citep{metallicity_gradient_2,Kovtyukh:2022} and the interstellar medium \citep{DeCia:2021}, amongst others. With the advent of \Gaia\ Data Release (DR) 3 \citep{GaiaCollVallenari:2023}, azimuthal variations were mapped in the metallicity of both young and old stars \citep{spiral_arm_DR3,DR3_chemical_cartography,Hawkins_2023}. More precisely, using young giants ($\lessapprox$ 150 Myr), \cite{spiral_arm_DR3} mapped local metallicity enhancements at the level of $\sim 0.1$ dex, which appeared to be statistically correlated with the position of the nearest spiral arms in the Galaxy. Using giant stars in Gaia DR3, \cite{Hawkins_2023} detected azimuthal variations which appeared to be co-located with the spiral arms; however, using OBAF-type stars in the Large Sky Area Multi-Object Fibre Spectroscopic Telescope (LAMOST), they found that the correlation was not evident. Based on numerical simulations, spiral arms are expected to drive azimuthal variations in the disc metallicity \citep{Grand:2016,Khoperskov:2018,Khoperskov_2023, simulation_abundance_Jr}. Moreover, azimuthal metallicity substructures can be caused by the radial migration induced by satellite galaxies \citep{Carr_22} and/or the Galactic bar \citep{DiMatteo:2013,Filion_2023}.

This complexity reveals the need of further constraints on the structural and chemo-dynamical characteristics of the Galactic disc in both the radial and the azimuthal dimensions.  

Up to now, analytic chemical evolution models of the Galactic disc ignore azimuthal surface density variations, with the recent exception of the 2D \cite{2dsimulation_spitoni,2dsimulation_spitoni_2} or \cite{molla2019} models. One of the main results of \cite{2dsimulation_spitoni_2} 
is that elements synthesised on short time scales, like $\alpha$-elements, exhibit larger abundance fluctuations along the spiral arms than heavier elements like iron (Fe) and barium (Ba) produced with a larger time delay. The aim of this study is a detailed analysis of the chemical inhomogeneities in the MW disc, with the goal of detecting the spiral arm signatures in calcium and magnesium, and more generally to explore radial and azimuthal chemical variations in the disc.

In Section~\ref{data_selection} the details of our data selections and methodology are explained. Section~\ref{results} describes our results and Section \ref{Conclusion} summarizes our discussions and conclusions.

\section{Data selection and methodology}
\label{data_selection}
\label{method}

To select tracers of disc stellar populations and map large-scale inhomogeneities in individual chemical abundances, we followed the procedure described in \cite{spiral_arm_DR3} (hereafter referred to as Paper I). We used stellar atmospheric parameters (effective temperature $\Teff$, surface gravity $log(g)$, global metallicity \mh), calcium \cafe~and magnesium \mgfe~abundances derived from $\Gaia$~RVS spectra by the General Stellar Parametriser-Spectroscopy ($\gspspec$\footnote{In the \gspspec~catalogue, \mh\ traces the iron abundances \feh\ as iron lines dominate the non-$\alpha$ element feature used by \gspspec. The \alphafe~parameter is dominated by the \cafe~abundance, since the Ca II triplet absorption dominates the $\alpha$-element signatures in the Radial Velocity Spectrometer (RVS) spectral domain.}) module \citep[][]{DR3_RVS}. Each selected element abundance (except the metallicity), is calibrated as a function of $\Teff$ using the polynomials from the end of Table~4 of \cite{DR3_RVS}, lately extended in Table~A.1 of \cite{DoubleRGBs} for the $log(g)$ and the magnesium ratio. In addition, we make use of the geometric distances from \cite{Bailer-Jones_distances} and the Galactic velocities, calculated in \cite{DR3_chemical_cartography}. We performed different quality selections in the astrophysical parameters, the astrometry, the distances and the kinematics to minimize the presence of possible contaminants (see more details in Appendix \ref{Appendix_dataselection}).

\begin{figure}[]
    \centering
    \begin{subfigure}{\linewidth}
        \centering
        \includegraphics[width=1\textwidth]{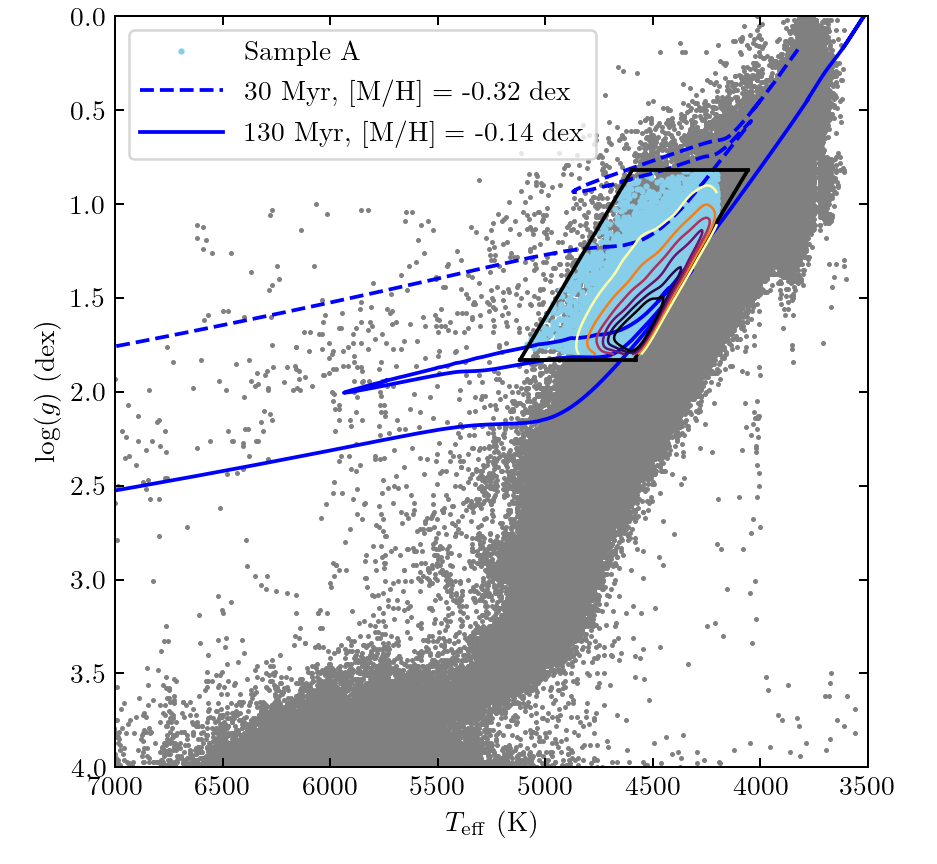}
    \end{subfigure}
    \begin{subfigure}{\linewidth}
        \centering
        \includegraphics[width=1\textwidth]{{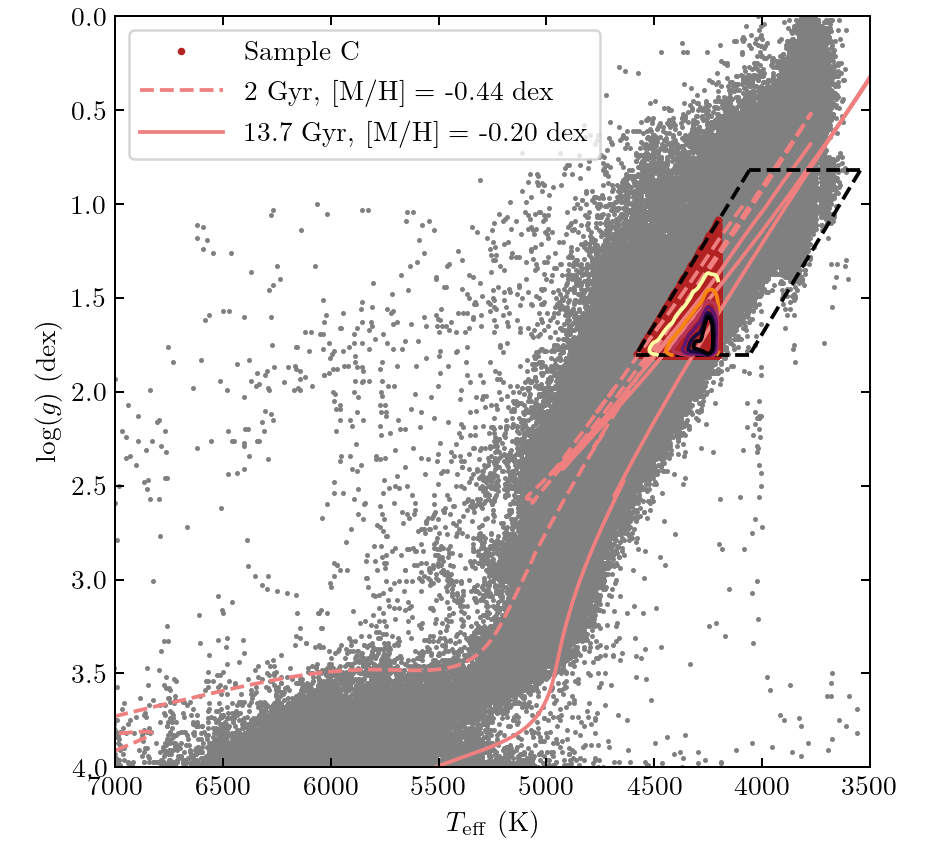}}
    \end{subfigure}
    \caption{\textit{Upper panel}: Selection of sample A targets (blue dots) in the Kiel diagram using the selection criteria of the calcium query (see Appendix \ref{Appendix_dataselection}). The box in solid lines are the initial sample A before applying the cutoff in $\Teff$. As a visual reference, the grey points represent the MW population from the initial selection (including all stellar types from the \gspspec\ catalogue). We overplotted isochrones based on BaSTI stellar evolution models, and density contours of the stars distribution with contour lines enclosing fractions of 90, 75, 60, 45, 30, and 20\% of the total number of stars. \textit{Lower panel}: Same figure showing the sample C selection (red dots). The box in dashed lines are the initial sample C before applying the cutoff in $\Teff$.}
    \label{kiel_diagram}
\end{figure}

From the above-mentioned preliminary selection, we are particularly interested in selecting giant stars, since they allow us to sample a relatively large volume of the Galactic disc, and to perform a robust statistical analysis thanks to the large number of high-quality chemical measurements. Specifically, we aim to select two samples of typically young (relatively hotter) and old (relatively cooler) giants stars, hereafter labeled as samples A and C, respectively. Following Paper I, we select stars in the Kiel diagram using cuts in $\Teff$ and log($g$), which create inclined boxes as shown in Fig. \ref{kiel_diagram}, within the following limits :\\

\hspace{-0.5cm}\verb| log(g)_{uncalibrated} < 1.5 & |\\
\verb| log(g)_{uncalib} > 0.5 &|
\verb|(log(g)_{uncalib} > (coeff*teff + interc_left))&|
\verb|(log(g)_{uncalib} < (coeff*teff + interc_right))&|
\verb| abs(Z) < 0.75 kpc| \\
\\
where the adopted slope \verb|coeff|, and the intercepts delimit the selected regions for samples~A and C (cf. Paper~I). The surface gravity used here is the uncalibrated one, and $Z$ is the distance to the Galactic plane. Compared to Paper I, we simply included greater constraints on \mh~uncertainties, adding \alphafe~uncertainties, individual abundance flags and considered calibrated parameters. In order to validate the robustness of the selected abundance statistics, we fixed a condition where the \xfe~uncertainty has to be twice higher than the standard deviation of the studied chemical species, $\sigma_{\xfe}$.


We also imposed {an additional} temperature constraint, requiring $\Teff>$ 4200~K. As shown in Fig. \ref{kiel_diagram}, this cut is almost irrelevant for sample A, but is very important for sample C (as it removes 211524 stars). 
The $\Teff>$ 4200~K criteria aims to minimize selection function effects, as explained in the following. The scanning law of the \Gaia\ satellite leaves important signatures in the \alphafe\ abundance distribution of \Gaia\ \gspspec\ catalog, as shown in Fig.~3 of \cite{DR3_chemical_cartography}. After performing some tests, we found that the scanning law artifacts are particularly evident at high Galactic latitudes \citep[see also][]{CantatGaudin:2024}, and quite evident for cooler stars (while hot stars tend to be more confined to the Galactic plane). By applying this $\Teff$ cut-off, we also limit the contribution of cooler stars, which are more difficult to parameterize. It helps to reduce the scanning law effects impacting sample C $\alpha$-abundances of this study composed of cooler stars. 

The process to obtain the initial queries and the final samples is explained in Appendix \ref{Appendix_dataselection}. In the end, sample A is preferentially populated by young stars (cold Blue Loop stars) tracing the spiral arms, while sample~C contains older disc populations (Red Giant Branch (RGB) stars). 
In this paper, we construct chemical maps based on the two samples selected in this section, with the goal of comparing two stellar populations of different age.

\begin{figure*}[htbp]
\begin{subfigure}{0.5\textwidth}
\includegraphics[width=1\linewidth]{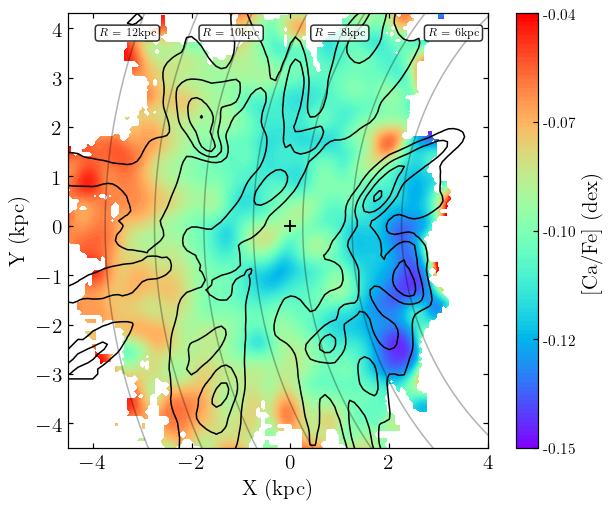} 
\label{XYmap_A_cafe}
\end{subfigure}
\begin{subfigure}{0.5\textwidth}
\includegraphics[width=1\linewidth]{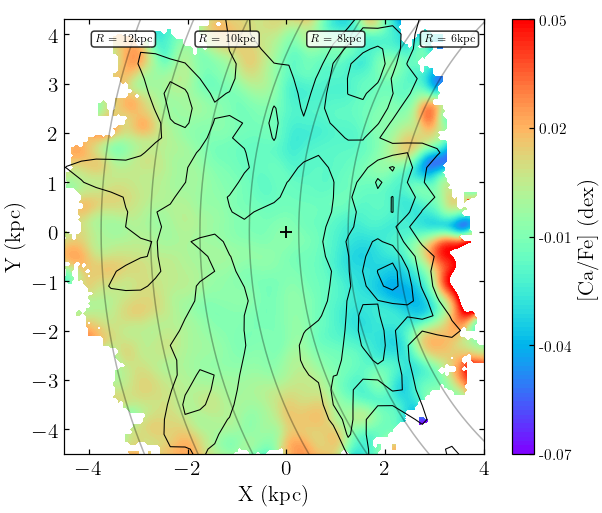}
\label{XYmap_C_cafe}
\end{subfigure}
    \caption{Chemical inhomogeneities in the Galactic disc for two different samples. \textit{Left panel:} Map of \cafe\ abundances for sample A. Black contours indicate the position of the spiral arms obtained using Upper Main Sequence (UMS) stars from \cite{spiral_arm_DR3} and trace from left to right, the Perseus arm, the Local arm, and the Sagittarius-Carina arm. The position of the Sun is shown by the black cross at origin. The Galactic centre is to the right and the Galactic rotation is going clockwise. Rings of constant Galactocentric radius are overplotted as grey solid lines. \textit{Right panel:} Same as the left panel, but for sample C and using the position of the spiral arms found in the subsample of giant stars in \cite{spiral_arm_radial_action} as black contours. A weak signature of the \Gaia\ scanning law mentioned in the selection function (see Appendix \ref{Appendix_dataselection}) is still visible, particularly around X $\sim$ 0 kpc and Y = (-4.5, -1) kpc or Y = (2, 4.5) kpc.}
\label{XYmap_cafe}
\end{figure*}

\begin{figure*}[htbp]
\begin{subfigure}{0.5\textwidth}
\includegraphics[width=1\linewidth]{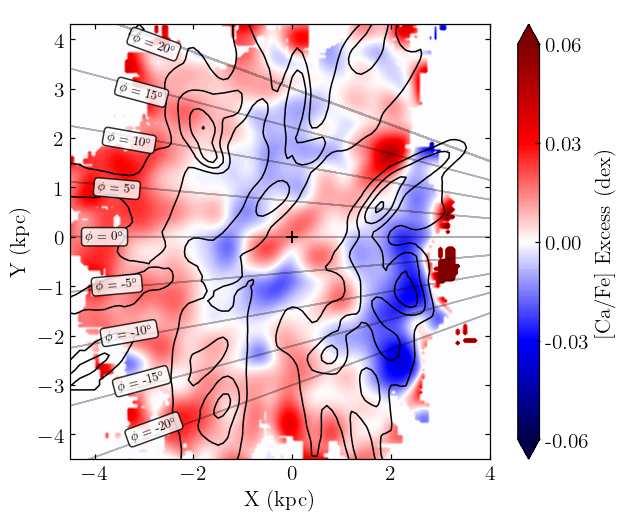} 
\end{subfigure}
\begin{subfigure}{0.5\textwidth}
\includegraphics[width=1\linewidth]{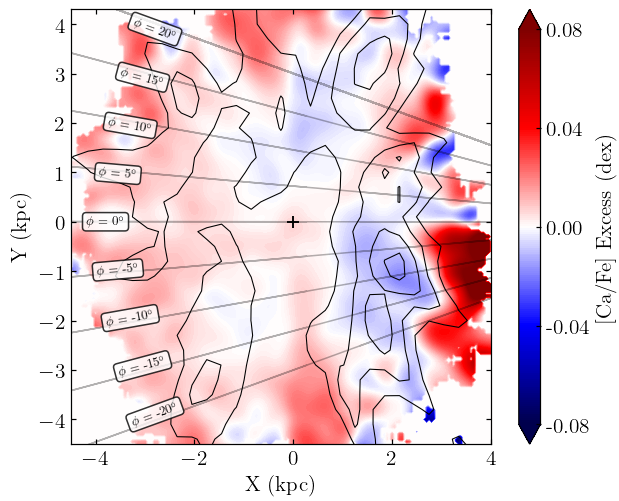}
\end{subfigure}
    \caption{Same as Fig. \ref{XYmap_cafe}, but now showing the maps of \cafe~excess and overplotting lines at different azimuthal angles in grey. \textit{Left panel:} Mean \cafe~excess in heliocentric coordinates for sample A. \textit{Right panel:} Mean \cafe~excess in heliocentric coordinates for Sample C. }
\label{XYmap_cafeexcess}
\end{figure*}

Fig. \ref{kiel_diagram} shows the final selected area defining sample A (blue dots) and sample C (red dots) with the initial selection (grey dots) as a visual reference. Sample~A includes 11678 stars with \cafe~abundances, reaching distances up to $\sim$ 4 kpc around the Sun and presenting a median G magnitude of 11.40~mag (with 10.35~mag and 12.09~mag being the 1st and the 3rd quartiles of the distribution, respectively). On the other hand, sample C includes 74740 stars with \cafe\ abundances. It reaches distances up to $\sim$ 4 kpc around the Sun. The G magnitude distribution spans from 11.40~mag (first quartile) to 10.44~mag (third quartile), with a median value of 12.06~mag. In addition, sample A contains about 689 stars with \mgfe\ abundances covering a smaller volume of the disc ($\sim$ 1.5~kpc around the Sun). For sample C, the condition fixed to have \mgfe~uncertainty twice higher than the $\sigma_{\mgfe}$ was not achieved in the magnesium abundance. For this reason, in this paper we will not show the \mgfe~abundance for this sample. The different characteristics of samples A and C are summarised in Table \ref{data_selection_details_overfe} where uncertainties of all parameters are defined as half of the difference between its upper and lower confidence levels, e.g. \verb|Teff|$_{\verb|unc|}$ = [\verb|Teff|$_{\verb|upper|}$-\verb|Teff|$_{\verb|lower|}$]/2 \citep{DR3_RVS}. We also checked the completeness of our dataset comparing the spectral type classification\footnote{Classification done using both RVS and $BP/RP$ spectra to select hot stars for the parametrization} of the Extended Stellar Parametrizer (ESP) catalogue \citep{ESP_cat} against \gspspec, where the former is more complete to the faint magnitude range as it uses \Gaia\ $BP/RP$ spectra. With respect to the ESP catalogue, we obtained a completeness of 93\% down to $G < 12$ mag for the entire query, a completeness of 84\% and 87\% down to $G < 12$ mag for samples A and C respectively.


In addition, as presented in Fig.~\ref{kiel_diagram}, we have estimated the typical age range within the selected portion of the Kiel diagram using BaSTI\footnote{\url{http://basti-iac.oa-abruzzo.inaf.it/isocs.html}}$^{,}$\footnote{The selected isochrones are defined by : Solar-Scaled \alphafe\ = 0.0, Overshooting:Yes, Diffusion:Yes, Mass loss:n=0.3, He=0.247} stellar isochrones \citep{Basti}. Indeed, it has been shown that high precision \gspspec\ data allow to break the age-metallicity degeneracy along the Giant Branch \citep{DoubleRGBs}. To take into account of the age-metallicity degeneracy, the value of the first quartile of the metallicity distribution ($\mh=-0.32\pm0.015$ dex) has been used for the hottest isochrone, and its third quartile ($\mh=-0.14\pm0.035$~dex) for the coolest one in the case of the youngest sample. Hence, varying the BaSTI age parameter to include the 90\% density of our sample leads to an approximate age range between 30 and 130~Myr. Following a similar approach, we found that sample C contains stars older than 2 Gyr, with a first quartile of the metallicity distribution equal to $\mh=-0.44\pm0.015$~dex for the hottest isochrone, and a third quartile of $\mh=-0.20\pm0.035$~dex for the coolest one. We also used stellar kinematics as another proxy for the typical age of the samples. Fig. \ref{hist_vphi} represents the azimuthal velocities dispersions V$_{\phi}$. The Galactic azimuthal velocities for sample A is more peaked than the original sample C stars (as for the subsets of sample C sorted by age), as expected for cooler (i.e. younger) populations. Moreover, as we can see, older the population, broader the distribution, followed by a lower V$_{\phi}$ median values. Therefore, the distributions presented in Fig. \ref{hist_vphi} confirm that sample C is dominated by stars typically older than sample A.

\section{Abundances maps for young and old disc populations}
\label{results}

\subsection{Calcium maps}
\label{results_maps}


The left panel of Fig.~\ref{XYmap_cafe} presents the map of the mean calcium abundance with respect to iron $\langle\rm\cafe\rangle$ in the Galactic plane, using a smoothing Gaussian bivariate kernel\footnote{As discussed in \citep{spiral_arm_EDR3}, the choice of kernel type (e.g. Gaussian, Epanechnikov...) doesn't have a significant impact on the obtained maps.} with a local bandwidth, $h$, of 240 pc (see Appendix C of Paper~I for details about the Gaussian bivariate kernel). For comparison, we also overplot as solid black lines the segments of the nearest spiral arms as traced by the density distribution of Upper Main Sequence\footnote{Revised selection of more than 600000 UMS stars from a cross match between photometric measurements from the 2-Micron All Sky Survey (2-MASS) and \Gaia~\cite[cf. Sect. 2.1 of][]{spiral_arm_EDR3}.} (UMS) stars \cite{spiral_arm_EDR3}. The solid contours show from left to right, the Perseus arm, the Local arm, and the Sagittarius-Carina (Sag-Car) arm. As expected, a \cafe\ radial gradient with more Ca-poor towards the inner parts of the disc is visible, in agreement with the literature \citep{alpha_radial_gradient_2, alpha_radial_gradient, alpha_radial_gradient_3}. In addition, we detect for the first time strong azimuthal fluctuations traced by the \cafe\ abundance.

The right panel of Fig.~\ref{XYmap_cafe}, shows the same results for sample~C with $h$=200 pc. The solid black lines show the overdensity contours of the disc giant star population\footnote{Sample selected from \Gaia~DR3 using photometry and imposing criteria in order to limit contribution of main sequence and faint dwarf stars \citep[cf. Appendix C][]{spiral_arm_radial_action}.} studied by \cite{spiral_arm_radial_action} showing also from left to right, the Perseus arm, the Local arm, and the Sag-Car arm. Although the \cite{spiral_arm_radial_action} selection is based on photometric considerations and could be partially contaminated by young stars, their kinematical analysis (c.f. their Appendix~C) suggests that old disc stars dominate the dataset. It is worthwhile mentioning that an older age margin (> 3~Gyr) in the Kiel diagram selection for sample C (further reducing the possible contamination from younger stars) has no visible impact on the abundance maps. In addition, a \cafe\ radial gradient with more Ca-poor towards the inner parts of the disc is again visible. Also, we detected for the first time azimuthal fluctuations traced by the \cafe\ abundance even if their are less evident. The two maps illustrate that a simple radial \cafe\ abundance gradient is not enough to explain the observed abundances distribution. Abundance inhomogeneities appear at different radii and for different azimuths.


To improve the mapping of the Ca abundance inhomogeneities, we define the \cafe\ excess as $\rm\cafe_{loc} - \cafe_{large}$, where $\rm\cafe_{loc}$ and $\rm\cafe_{large}$ represent the mean \cafe\ smoothed on a local or large scale, respectively. The large scale corresponds to a bandwidth 6 times larger the local one (i.e. $h_{large} = 1440$ pc for sample A and $h_{large} = 1200$ pc for sample C). The resulting map of the \cafe\ excess is presented on the left and right panels of Fig.~\ref{XYmap_cafeexcess} for sample A and sample C, respectively. In the left panel, the \cafe\ abundance does not increase monotonically as a function of Galactocentric radii R (from right to left). Indeed, it presents local decreases in \cafe\ in correspondence of the Local and the Sag-Car arms being more \cafe\ poor than the inter-arm regions. The obtained \cafe\ patterns can be compared to the \mh\ ones presented in left panel of Fig. \ref{xymaps_mh_excess} (or see Paper~I for comparison), which showed that stars inside the spiral arms tend to be more metal-rich than those in the inter-arm regions. The anti-correlation between the \mh\ and \cafe\ distributions is characterised by a Spearman coefficient \citep{Spearman04} of -0.63, corresponding to a strong anticorrelation \citep{spearman_coeff}. It is important to note that the \mh\ and the \cafe\ are derived separately by the \gspspec\ module using different reference grids and methodologies. Radial and azimuthal fluctuations are clear but the magnitude of the observed abundance fluctuations is not constant along a given spiral arm. Indeed, the Perseus arm and the lower part of the Local Arm show no \cafe\ deficiency and, on the contrary the region around (X, Y) = (0, -1) kpc shows a negative \cafe\ variation that does not correspond to the UMS overdensity contours. About the right panel presenting the excess map of sample C, it is interesting to point out the sharp correspondence between \cite{spiral_arm_radial_action} old disc overdensity contours and our observed features. Again, a \cafe\ deficiency in the Sag-Car arm and for the Local Arm area are highlighted. To quantify the correlation between the \mh\ and \cafe\ distributions we estimated the Spearman coefficient, compared to sample A, the coefficient is equal to -0.68, corresponding again to a strong negative correlation. As for the young sample~A, the \cafe\ maps do not clearly match the stellar density distribution for the entire Perseus Arm or for the Local Arm at negative azimuths. 

\begin{figure*}[htbp]
\begin{subfigure}{0.5\textwidth}
\includegraphics[width=1\linewidth]{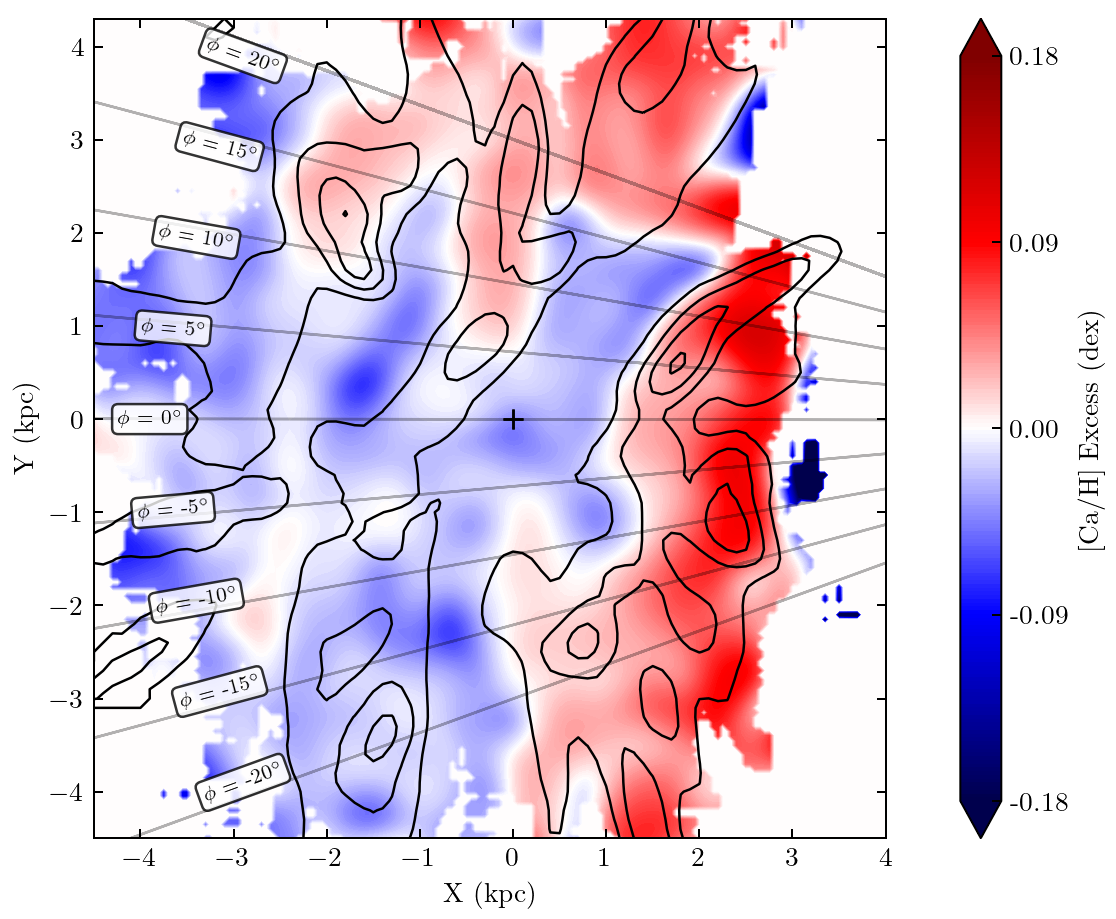}
\end{subfigure}
\begin{subfigure}{0.5\textwidth}
\includegraphics[width=1\linewidth]{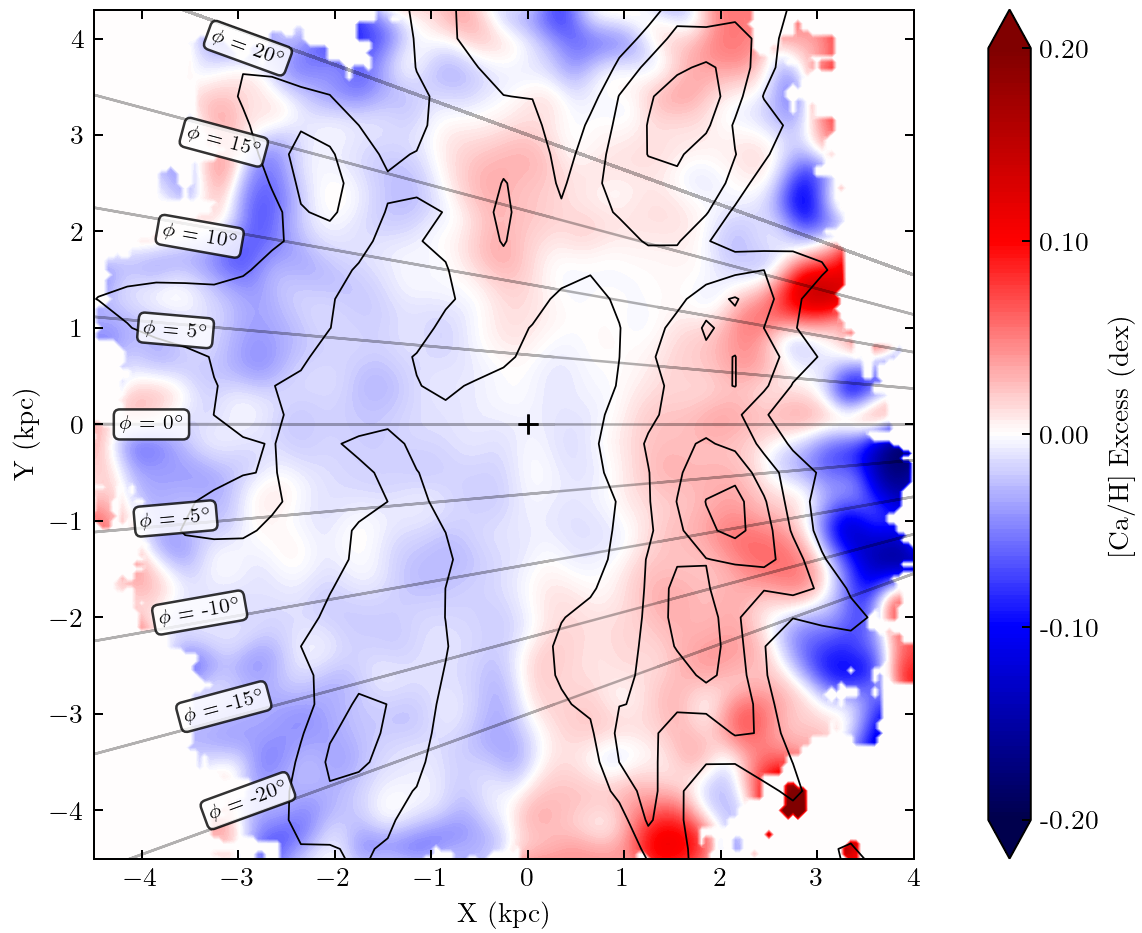}
\end{subfigure}
\caption{\textit{Left panel:} Map of \cah~excess for sample A, including the spiral arms contours from \cite{spiral_arm_EDR3} adopting the same local bandwidth as for \cafe. \textit{Right panel:} Same for sample C, including the spiral arms contours from \cite{spiral_arm_radial_action} adopting the same local bandwidth as for \cafe.}
\label{xymaps_cah_excess}
\end{figure*}

\begin{figure*}[htbp]
\begin{subfigure}{0.5\textwidth}
\includegraphics[width=1\linewidth]{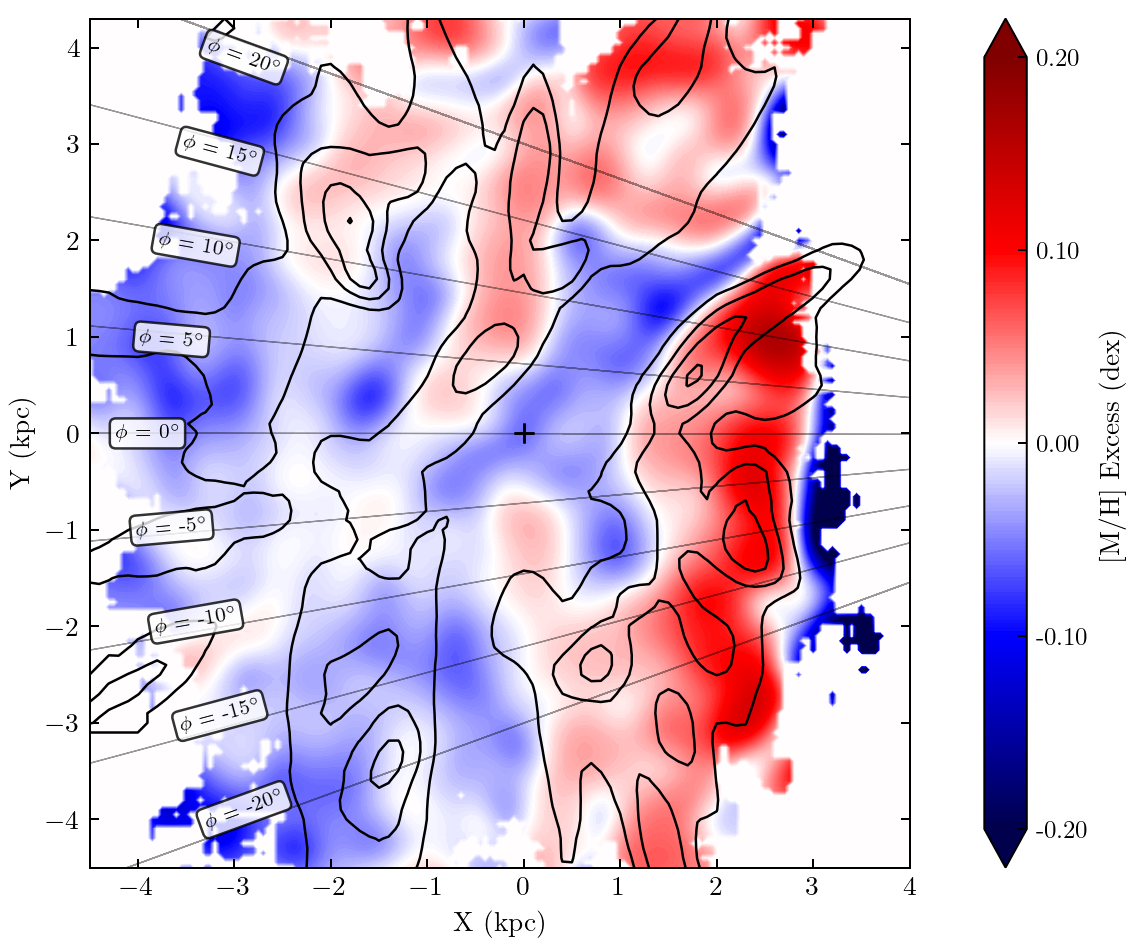}
\end{subfigure}
\begin{subfigure}{0.5\textwidth}
\includegraphics[width=1\linewidth]{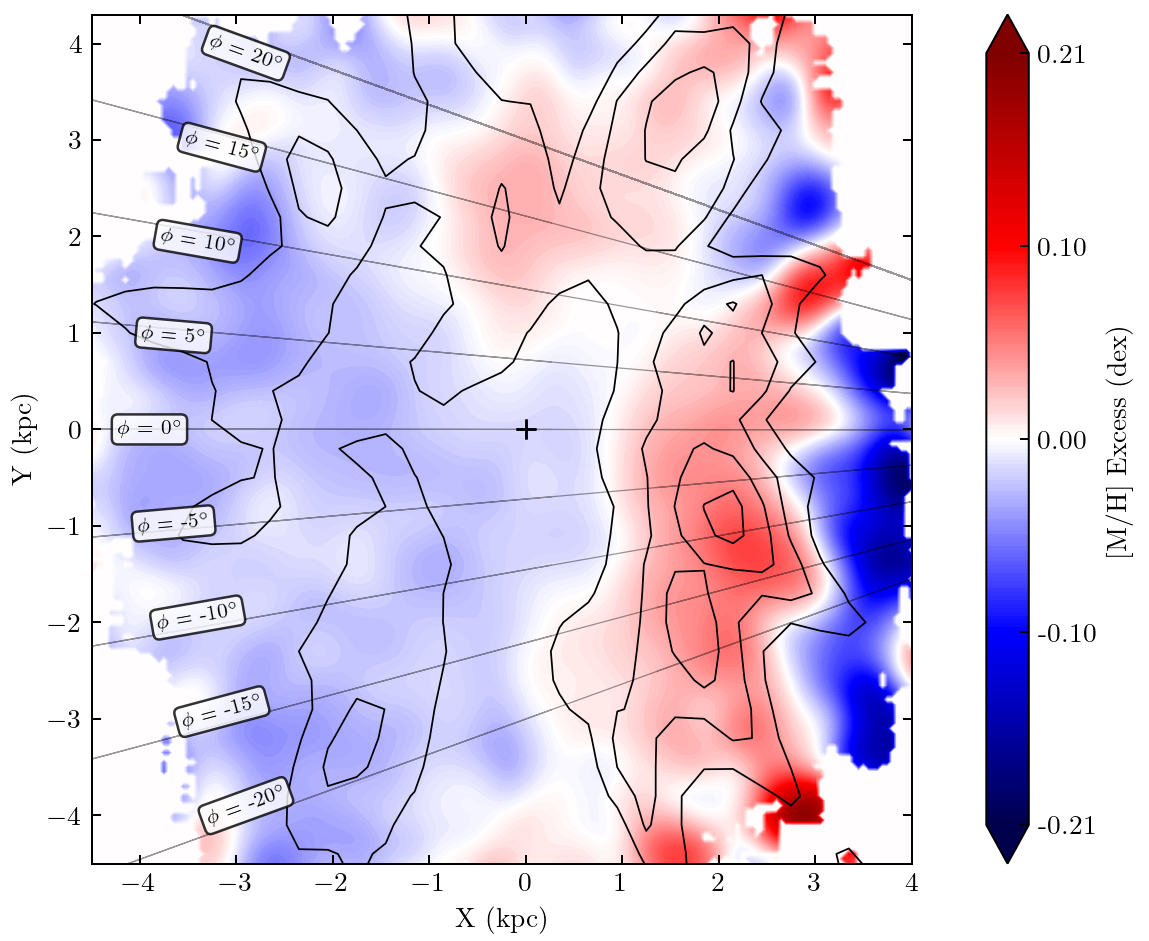}
\end{subfigure}
\caption{\textit{Left panel:} Map of \mh~excess for sample A, including the spiral arms contours from \cite{spiral_arm_EDR3} and adopting a $h_{local}$ = 200 pc and a larger scale length, $h$, 6 times higher. \textit{Right panel:} Same for sample C, including the spiral arms contours from \cite{spiral_arm_radial_action} and adopting a $h_{local}$ = 200 pc and a $h_{large}$ = 1200 pc (i.e. 6 times higher).}
\label{xymaps_mh_excess}
\end{figure*}

It is also possible to analyze separately the calcium abundances and the iron ones (i.e. the metallicity). To this purpose, the \cafe\ and the \mh\ abundances are combined to obtain the \cah \footnote{\cah = \cafe+\mh} excess map presented in Fig.~\ref{xymaps_cah_excess}. In our datasets, \mh\ and \cah\ median uncertainties are typically twice smaller than the typical global dispersion of the sample (see Table \ref{data_selection_details_overfe} for the statistical properties of our samples for different chemical abundances). For both young and old populations, the maps of Fig. \ref{xymaps_cah_excess} allow us to directly trace the chemical evolution of calcium, separately from iron. Similarly to metallicity (cf. Fig. \ref{xymaps_mh_excess}), the calcium abundances show positive fluctuations for the Sagittarius-Carina (Sag-Car) arm and the upper part ($Y > -1$ kpc, I and II Galactic quadrants) of the Local arm reinforcing the view of a faster chemical evolution in the arms than in the inter-arm regions. Those likeness between the \mh\ and \cah\ maps are characterised by a Spearman coefficient equal to $0.96$ both for samples A and C, corresponding to a very strong positive correlation. Again the radial and azimuthal fluctuations are evident, noting that the magnitude of the observed abundance fluctuations is not constant along a given spiral arm. In the Local arm, similarly to \cafe\ excess maps for both samples, regions at $Y<0$ are significantly more \rm\cah-poor or \rm\mh-poor than those at $Y>0$.
About the Perseus arm, the spiral arms signatures are only obtained for the \cah\ or \mh\ at positive azimuths (larger than 10\deg, in the Cassiopeia region) and only for sample A. Again only for the young population disc, we retrieve the region around (X, Y) = (0, -1) kpc showing this time a positive \cah\ and \mh\ fluctuations that do not correspond to the UMS overdensity contours. In Section \ref{Conclusion}, we will discuss about the plausible explanations proposed for understanding the differences between the density contours and obtained fluctuations, and the variation of abundances within the spiral arms that are not constant along a given arm.


\subsection{Quantification of the observed abundance variations}
\label{results_quantification}

\begin{figure*}[htbp]
\begin{subfigure}{0.5\textwidth}
\includegraphics[width=1\linewidth]
{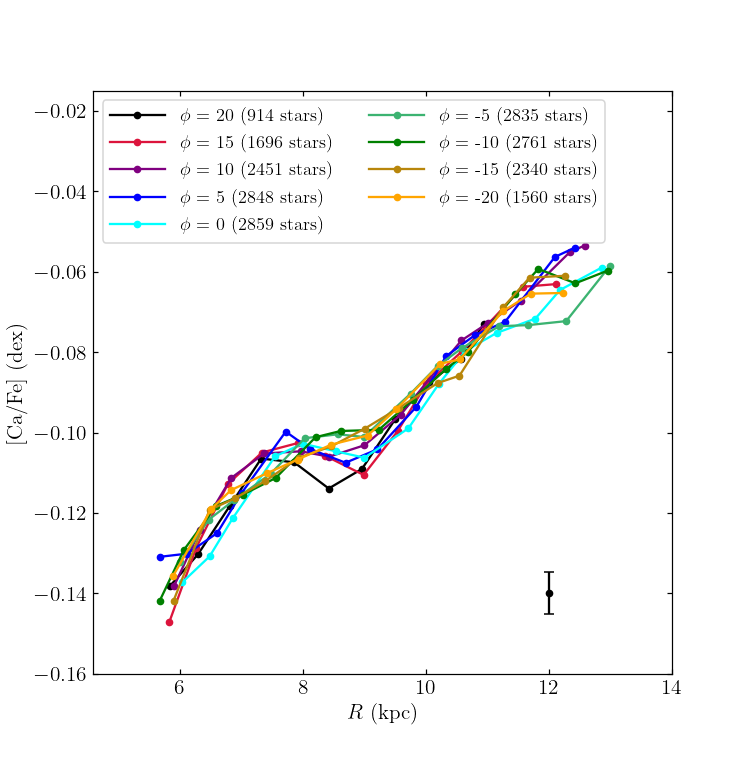}
\end{subfigure}
\begin{subfigure}{0.5\textwidth}
\includegraphics[width=1\linewidth]
{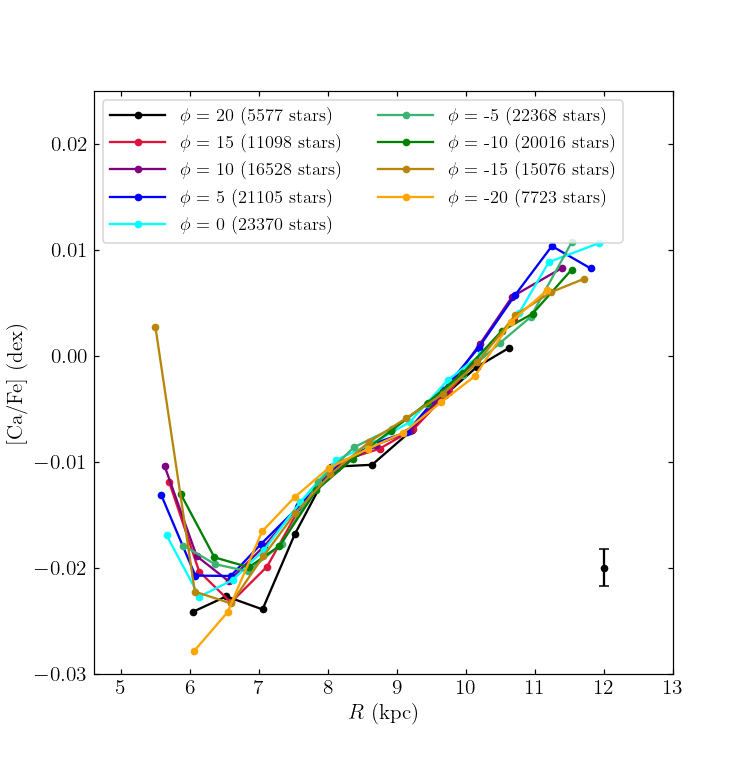}
\end{subfigure}
\caption{\textit{Left panel}: Radial gradients in \cafe\ abundances of sample A. The colour code illustrates the results of the running mean applied on the stars contained in the different bins of 5$\deg$. Each dot corresponds to the mean of all stars every 1.1 kpc for a specific bins of 5$\deg$. \textit{Right panel}: Same as left panel for sample C.}
\label{radial_cafe}
\end{figure*}

To better characterize the observed inhomogeneities in both samples, radial gradients in the \cafe\ abundances are shown in Fig. \ref{radial_cafe}. The left panel of Fig. \ref{radial_cafe} shows the \cafe\ radial gradient for sample A computed by using a running mean of 5$\deg$ bin width in azimuth (based on the lines of different azimuthal angles in grey from Fig. \ref{XYmap_cafeexcess}). As previously mentioned, on top of the general increase in \cafe\ with Galactic radii, the radial gradients present oscillations corresponding to the azimuthal fluctuations. These radial variations appear to be correlated with the spiral arms in some regions, as discussed above. 


\begin{table}[]
\centering
\caption{Slopes and intercepts estimated by a linear fit of the radial \cafe~gradients in different azimuthal bins (c.f~Fig. \ref{radial_cafe}).}
\begin{tabular}{c|cc|}
\cline{2-3}
                        & \multicolumn{2}{c|}{Sample C}                    \\ \hline
$\phi$ {[}deg{]}     & Slope {[}dex.kpc$^{-1}${]} & Intercept {(}dex{)} \\ \hline
20 $\pm$ 5  & 0.0061 $\pm$ 0.0005         & -0.063 $\pm$ 0.004  \\
15 $\pm$ 5  & 0.0061 $\pm$ 0.0004         & -0.062$ \pm$ 0.003  \\
10 $\pm$ 5  & 0.0061 $\pm$ 0.0003         & -0.061 $\pm$ 0.003  \\
5 $\pm$ 5   & 0.0060 $\pm$ 0.0003         & -0.060 $\pm$ 0.002  \\
0 $\pm$ 5   & 0.0058 $\pm$ 0.0002         & -0.059 $\pm$ 0.002  \\
-5 $\pm$ 5  & 0.0057 $\pm$ 0.0003         & -0.058 $\pm$ 0.003  \\
-10 $\pm$ 5 & 0.0059 $\pm$ 0.0002         & -0.060 $\pm$ 0.002  \\
-15 $\pm$ 5 & 0.0061 $\pm$ 0.0003         & -0.062 $\pm$ 0.003  \\
-20 $\pm$ 5 & 0.0056 $\pm$ 0.0005         & -0.058 $\pm$ 0.004 
\label{radial_cafe_table}
\end{tabular}
\tablefoot{Only stars between $6.5 \leq R \leq 11$ kpc are considered in order to avoid selection effects in the inner highly extincted regions, as well as low number statistic areas.}
\end{table}

For sample C, the radial \cafe~gradients (cf. right panels of Fig.~\ref{radial_cafe}) 
show significant fluctuations for R$<$6 kpc that should be considered with caution. Although they could be border effects, it is important to note that these fluctuations are observed consistently in both metallicity and calcium: enrichment of the interstellar medium similar to that expected outside the spiral arms (i.e., increase in \cafe\ or depletion in \mh). Nevertheless, the observed abundances are also consistent with a chemical bias due to the loss of more metal-rich and \cafe\ poor stars in the Galactic plane caused by the higher interstellar extinction in those internal regions. Compared to sample A, the slopes observed in sample C highlight smoother radial trends in all azimuthal directions. The \cafe\ gradients vary gradually with $\phi$, becoming slightly steeper for $\phi >$ 0$^\circ$ (c.f. Table~\ref{radial_cafe_table}), and therefore, being anticorrelated with the \mh\ gradients reported in Paper~I for sample~C. However, some fluctuations due to azimuthal changes in the abundance ratios, although flatter than those observed for the young population, are still visible.

\begin{figure*}[htbp]
\begin{subfigure}{0.51\textwidth}
\includegraphics[width=1\linewidth]
{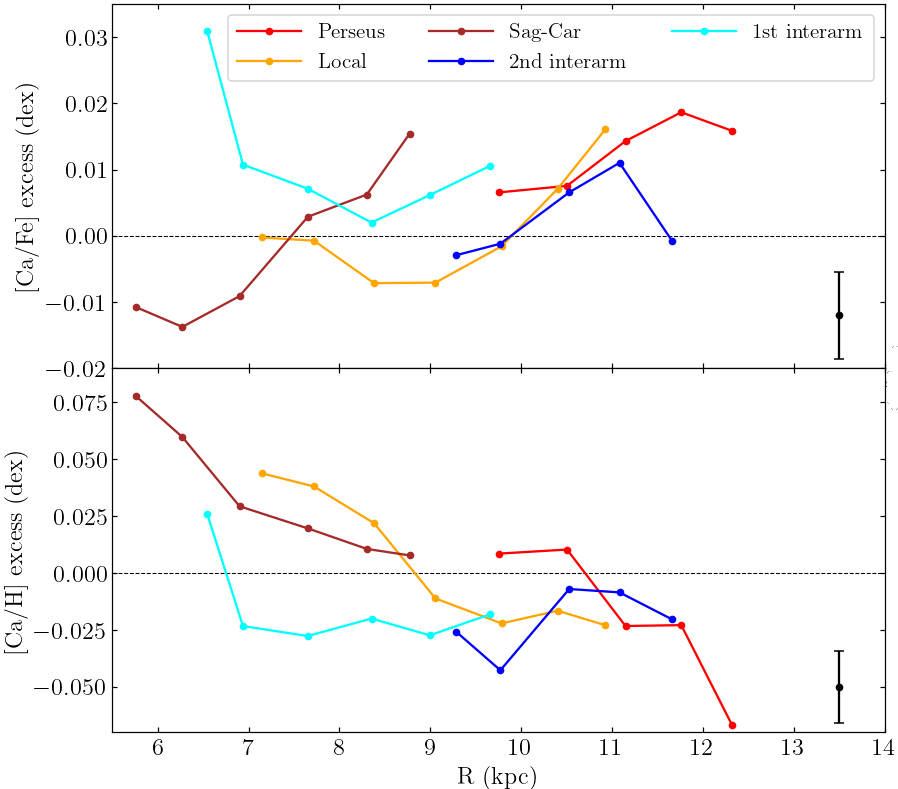} 
\end{subfigure}
\begin{subfigure}{0.50\textwidth}
\includegraphics[width=1\linewidth]{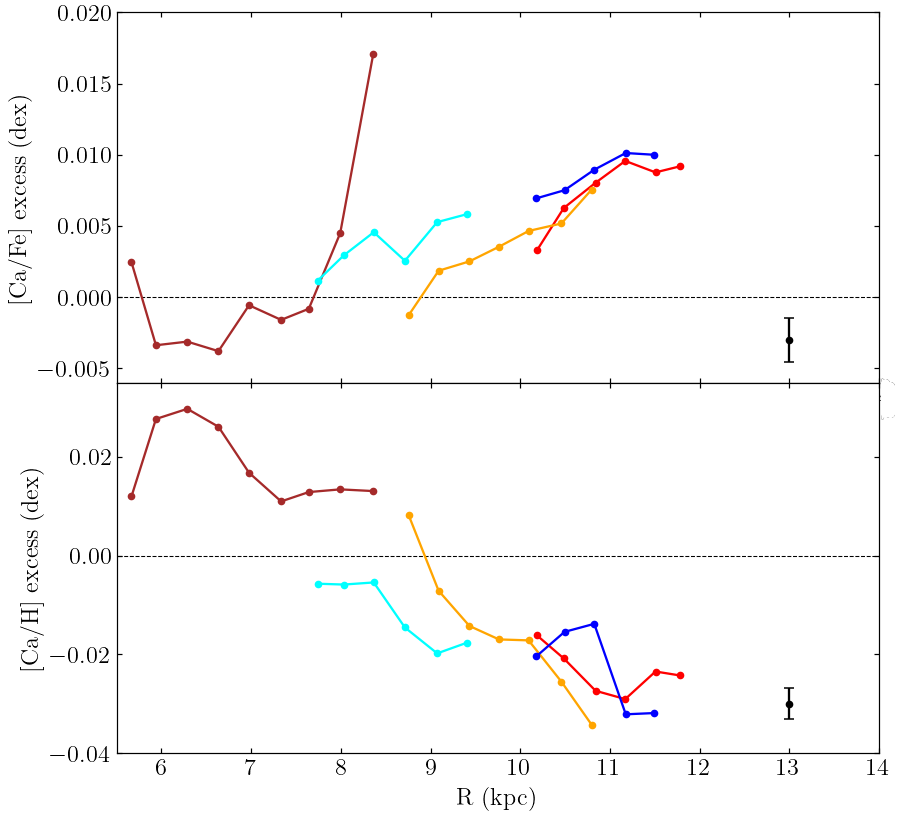} 
\end{subfigure}
    \caption{\textit{Left panels}: Azimuth-averaged radial distribution for sample A, selecting bins of 650 pc and excluding those with less than 20 stars. From the top to the bottom, excess of \cafe\ and \cah\ in an azimuth-averaged radial distribution are represented based on the currently known locations of the Galactic disc arms and interarms from the density distribution of the stars of the sample. 
    \textit{Right panels}: Same as left panels, but now for sample C, using bins of 350 pc in Galactocentric radius R. Bins with less than 200 stars were excluded.}
\label{radial_gradient_excess}
\end{figure*}

To better visualize the abundances fluctuations in the arm versus interarm regions, we selected in boxes the stars located inside or outside the spiral arms ovendensity in order to construct an azimuth-averaged radial distribution of each sample. Using the excess data, we computed the median excess variation of all stars within each azimuthal bin across the radial distribution. Fig \ref{radial_gradient_excess} shows the variation of the abundance excess as a function of Galactocentric radii for different regions in the Galactic disc. Positions in the $X-Y$ plane are assigned to a specific arm segment or to an interarm region based on the density distribution of stars. The first interarm corresponds to the area between the Local and Sag-Car arms, while the second interarm corresponds to the area between the Perseus and Local arms. This way of visualization indicates that not all the regions in the spiral arms exhibit the same chemical behaviour. 

Both for samples A and C, we retrieve broadly similar trends and variations of the chemical abundances. For the first panels with the \cafe\ excess, the interarms regions are generally \cafe\ richer compared compared to the arms regions, with the exception of the Perseus arm. In the \cah\ case, we have almost the same chemical pattern in the selected structures, that confirms the strong positive correlation found in Section \ref{results_maps}.
 For a given arm, the \cafe\ excess typically declines as a function of $R$ (and increased for \cah).

\subsection{Magnesium maps only for the youngest sample}
\label{results_mg}

\begin{figure}[htbp]
\includegraphics[width=1\linewidth]{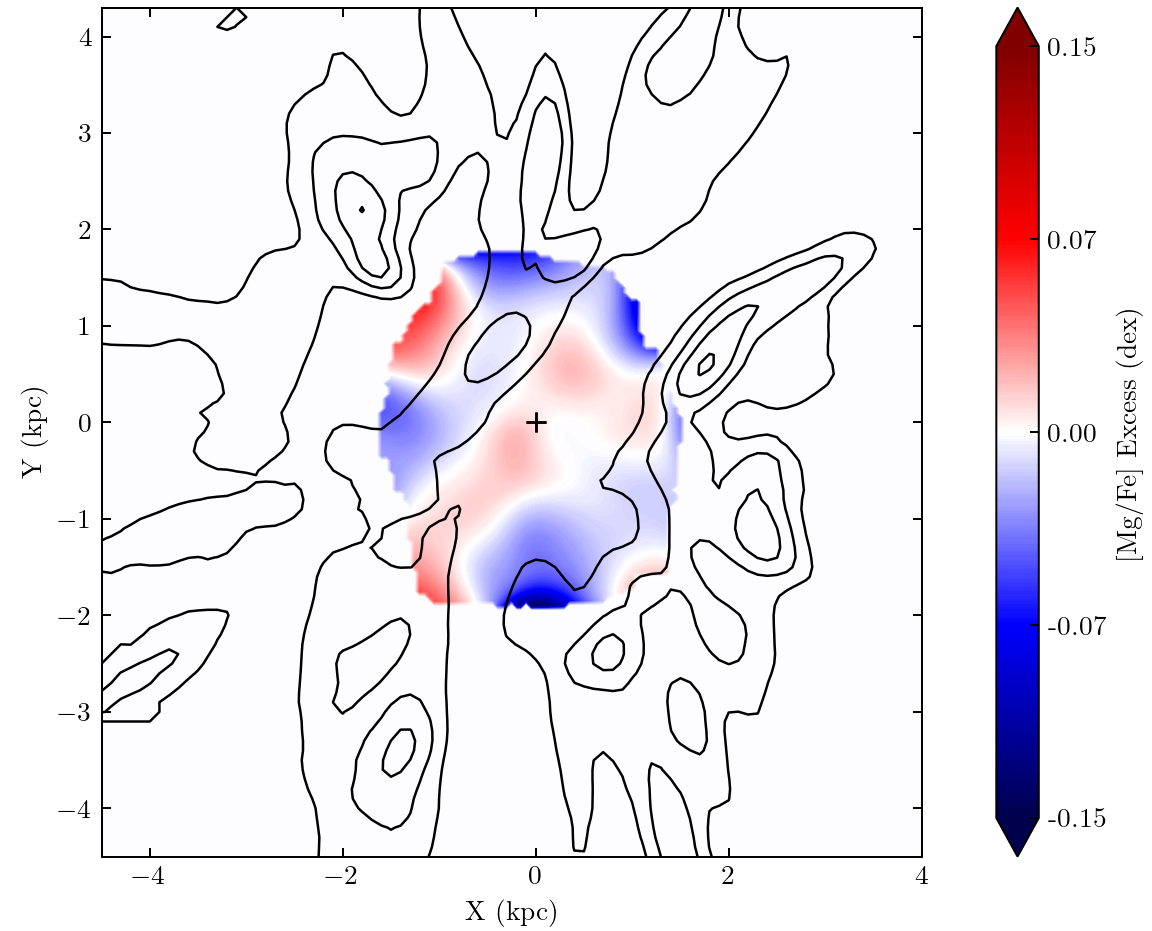}
\caption{\mgfe\ excess map for sample A overplotted by the spiral arms contours from \cite{spiral_arm_EDR3}. Here again, the local bandwidth is unchanged.}
\label{Xfe_excess_comparison_SampleA}
\end{figure}

The \cafe\ abundance maps can be compared to the \mgfe\ ones in a reduced area (reaching distances of up to about 1.2~kpc around the Sun), due to the imposed threshold in the signal-to-noise (SNR) (see Appendix \ref{Appendix_dataselection} for more details). Thus, sample~A includes only 689 stars with \mgfe~abundances which are characterised by a median G magnitude of 7.9~mag (with 7.1~mag and 8.3~mag being the 1st and the 3rd quartiles of the distribution, respectively). As for the case of \cafe, we computed the \mgfe\ excess to better map the abundances inhomogeneities from the difference \mgfe$_{loc}$ - \mgfe$_{large}$. Due to the shorter spatial coverage, the used large scale length for \mgfe$_{large}$ is 720 pc (i.e. 3 times larger than the one used for the local values \mgfe$_{loc}$ that is $h$ = 240 pc). Fig. \ref{Xfe_excess_comparison_SampleA} illustrates the resulting map of \mgfe\ excess. Despite the reduced area (and therefore the more important border effects), we observe patterns similar to those in \cafe\ for the Local Arm region and its surroundings: the stars within the Local Arm tend to be deficient in \mgfe\ and \cafe\ with respect to those in the inter-arm regions. At these distances, our number of stars is not enough to quantify the \mgfe\ fluctuations in the entire disc, but larger and deeper surveys, like \Gaia\ DR4, will allow their study as well as that of many other $\alpha$-elements.

\section{Discussion and conclusions}
\label{Conclusion}

\begin{figure*}[htbp]
\centering
\includegraphics[width=0.9\linewidth]{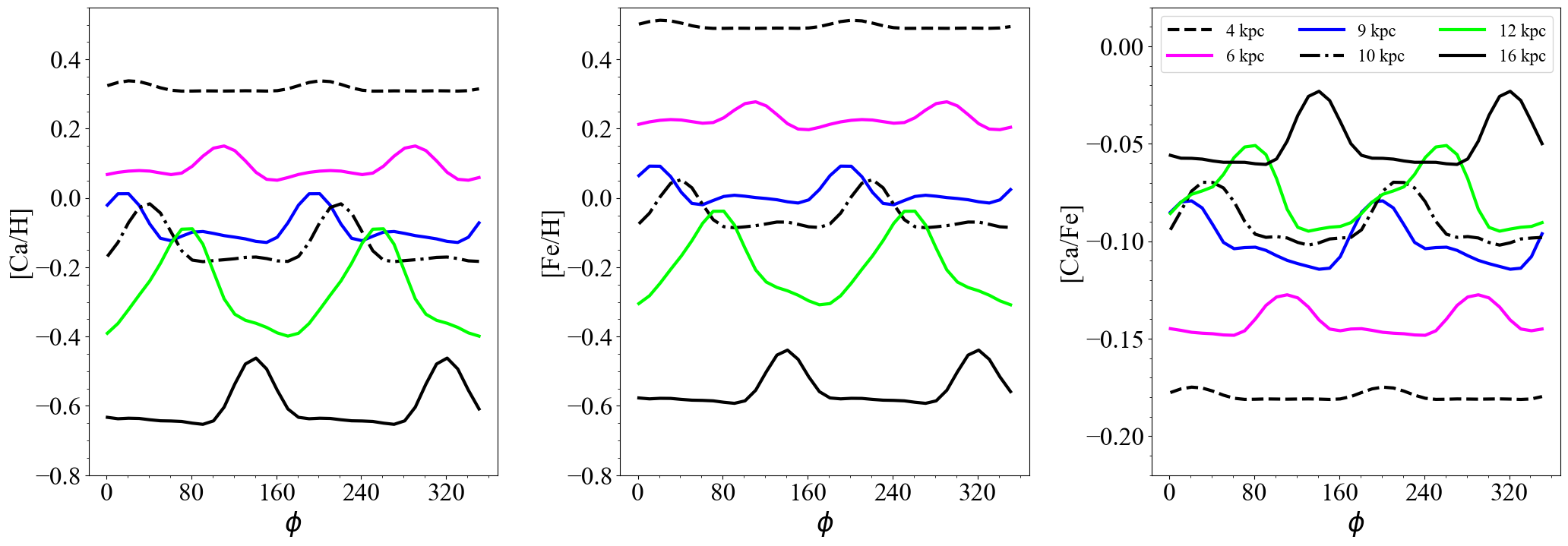}
\includegraphics[width=0.9\linewidth]{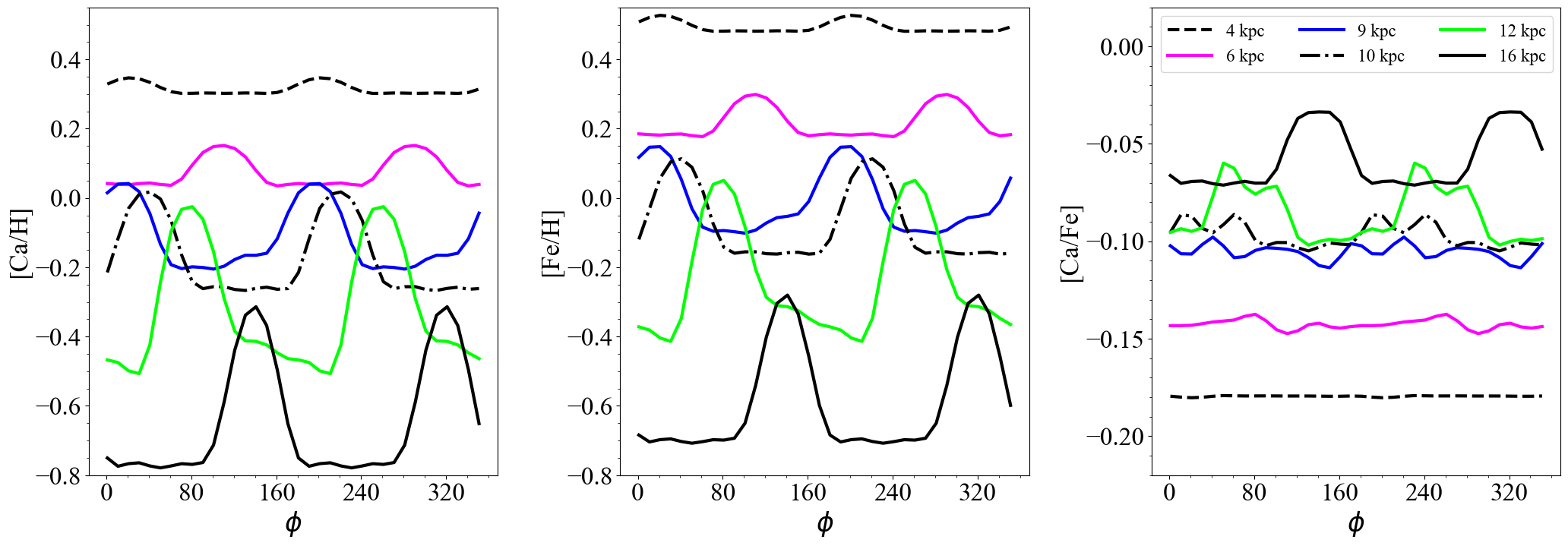}
\includegraphics[width=0.9\linewidth]{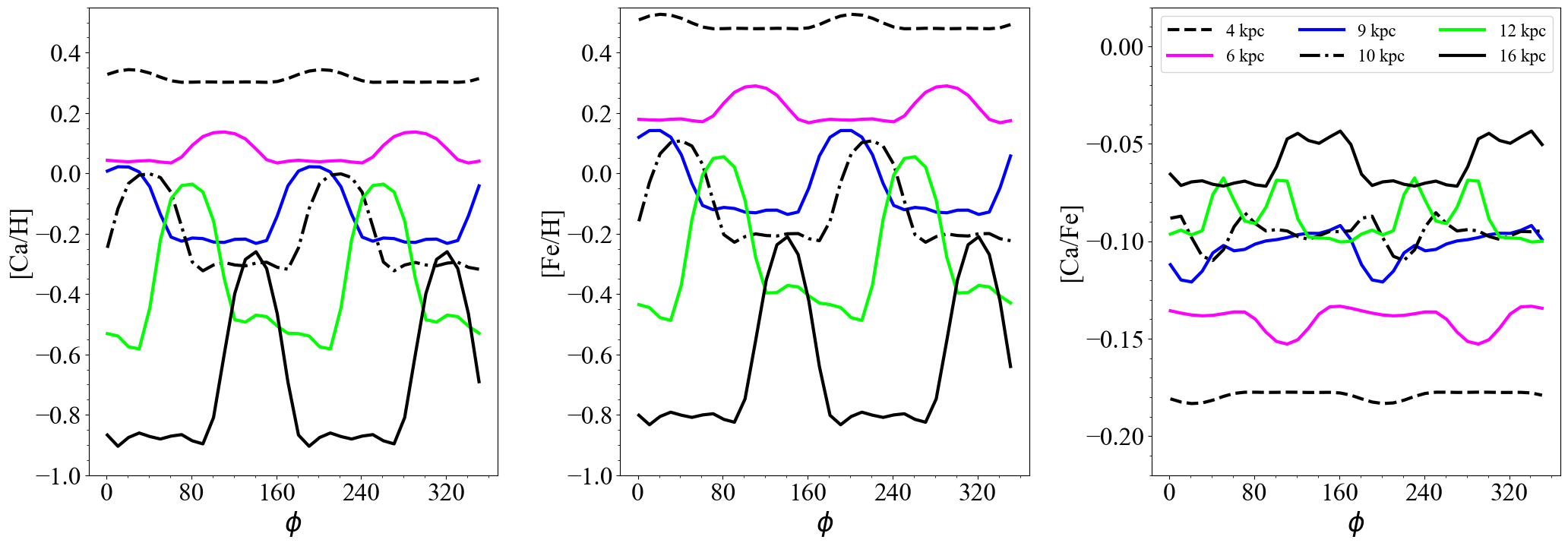}
\caption{Predicted present-day azimuthal variations in the [Ca/H] (first column), [Fe/H] (second column), and [Ca/Fe] (last column) ratios, respectively computed at different Galactocentric distances by the 2D chemical evolution model by  \citet{2dsimulation_spitoni_2} in presence of  multiple spiral structure. We  assume  that for the last 1 Gyr (first row, model A+C3 in \citealt{2dsimulation_spitoni_2}), 3 Gyr (second row) and 5 Gyr (third row), the co-rotation is extended at all distances. The coloured solid lines indicate the variations at the loci of co-rotation for the
three spiral structures characterised by different pattern speeds indicated in Fig. 1 of \citet{2dsimulation_spitoni_2} before extending them to all radii.}
\label{predicted_azimuthal_variation}
\end{figure*}

In this paper, we have produced 2D chemical abundance maps in the Galactic disc for two $\alpha$-elements: calcium and magnesium. These maps, based on the \cafe\ and \mgfe\ abundances derived by \Gaia\ \gspspec, show evidence of considerable radial and azimuthal inhomogeneities, which are accompanied by metallicity fluctuations. 
Our understanding of the disc chemical trends is changing from a simplistic 1D radial view, to a more complete 2D perspective combining radial trends, azimuthal tendencies, and small scale variations.

For young stellar populations, the inhomogeneities in $\alpha$-abundances are spatially coherent with the density contours of the Sagittarius-Carina arm, and the Local arm in the I and II Galactic quadrants \citep{spiral_arm_EDR3}, as already noted for the metallicity signatures in Paper~I or in figure \ref{xymaps_mh_excess} of this article. In the above-mentioned regions, the global emerging picture is that the stars within the spiral arms are globally i) more metal-rich (up to $\sim$0-0.20 dex, see also Paper I for comparison) and \cah-rich ($\sim$0.18 dex), ii) more \cafe-poor ($\sim$0.06 dex) and \mgfe-poor ($\sim$0.05 dex), than the stars in the inter-arms regions and iii) the magnitude of the observed fluctuations is not constant along a given spiral arm. 

Spiral arms are known possible drivers of chemical azimuthal variations, but their exact influence on the MW's disc is still unclear. This is in part due to our lack of knowledge of their dynamical nature \citep{Dobbs:2014}. \cite{LinShu:1964} described the spiral patterns in disc galaxies in terms of quasi-stationary density waves, moving with a single pattern speed \citep[see][ for a more recent review]{Shu:2016}. Following this framework, several works tried to estimate the pattern speed of the spiral arms in the MW, obtaining values ranging from $\sim$ 20 to 40 $\rm{km/s/kpc}$ \citep[see][ and references therein]{Gerhard:2011}. Using Open Clusters (OCs) in \Gaia\ Data Release 2, \cite{Dias:2019} obtained a common pattern speed for the Perseus, Local, and Sagittarius spiral Arms of $28.2 \pm 2.1~\rm{km/s/kpc}$, resulting in a corotation radius $R_C = 8.51 \pm 0.64~\rm{kpc}$, not far from the Sun's position. This is in contrast to the results obtained by \citet{Castro-Ginard:2021} based on OCs in \Gaia\ Early Data Release 3, who found that the spiral arms rotate at nearly the same speed as field stars at any given radius, discarding a common spiral pattern speed. Their results support the idea of transient spiral arms \citep{Toomre:1964,Sellwood:2012,Hunt:2018}. 

In literature, several works modeled spiral arms as a superposition of multiple spiral density waves having different pattern speeds \citep{Tagger:1987, Minchev:2012}. Following \cite{Minchev:2016}, \cite{2dsimulation_spitoni_2} modeled the spiral structure of our Galaxy as a combination of three different segments of varying pattern speeds, and developed a 2D chemical evolution model to test their impact on the distribution of chemical elements in the Galactic disc. 
In their model, the star formation is enhanced due to the passage of the spiral arms, hence, abundance fluctuations are produced with an amplitude depending on the considered chemical species. In the \cite{2dsimulation_spitoni_2} model, elements synthesised on short timescales (i.e. from short lifetime progenitors) exhibit larger abundance variations as they promptly report the effects of the spiral arms passage, contrary to other elements ejected into the interstellar medium (ISM) after a significant delay. On one hand, our analysis globally confirms the predicted Ca enhancement, although with important azimuthal variations. Ca is mainly produced by short lifetime progenitors (Type~II Super Novae (SN)), although long-live progenitors (Type~Ia SN) have a significant contribution\footnote{As indicated in \cite{kobayashi2020} or \cite{johnson_jenni_2020}, the synthesised fraction of Ca produced by Type Ia SN is around 45\%.}.

In this picture, as shown by \cite{2dsimulation_spitoni_2}, an important parameter is the time that a given region of the disc is affected by the spiral arms perturbations. This depends on the total life time of the spiral arms, but also on their pattern speed \citep{2dsimulation_spitoni_2,simulation_abundance_Jr}. The duration of the spiral arms influence on a particular region of the disc affects also the different abundance ratios, depending on the lifetime of the producers.

Our \cafe\ maps show a tendency to be \cafe-poor within the nearest spiral arms, indicating a higher abundance enhancement of Fe with respect to Ca. This suggests a more complex picture than predicted, as Fe has a larger contribution from long-lived progenitors (Type~Ia SN) than Ca, needing more time than $\alpha$-elements to dominate the chemical patterns. Consequently, in this context, the \cafe\ depletion of the spiral arms with respect to the inter-arm regions could result from a pattern speed close to or fluctuating around the disc co-rotation for a duration compatible with the predominance of Type Ia SN contribution to the spiral arms chemical pattern (i.e. the duration of co-rotation with the disc may be longer than previously thought). 

To test this hypothesis, we have run new simulations with the 2D chemical evolution model of \cite{2dsimulation_spitoni_2}.
Fig. \ref{predicted_azimuthal_variation} shows the azimuthal variations in \cah, \feh\ and \cafe\ abundances assuming a spiral arms co-rotation at all radii for 1, 3 and 5 Gyr, in the upper, middle and bottom panels respectively. Similarly to our observations (cf. left panel of Fig. \ref{XYmap_cafe}), only the models assuming a co-rotation of 3 and 5 Gyr, start to recover a deficiency in \cafe\ corresponding to the excess peaks in \cah\ and \feh\ variations (around 80 and 260\deg). 

This long co-rotation time between the spiral arms and the disc seems challenging from a dynamical point of view, although it is discussed by several authors in the literature. \cite{Sellwood_2014} presents evidence that a long recurrent spiral activity in disc galaxies can result from the superposition of a few transient spiral modes (cf. their Fig. 6). In their simulations, each mode lasts between 5 and 10 rotations at its co-rotation radius where its amplitude is the greatest. Following \cite{Minchev:2016}, \cite{2dsimulation_spitoni_2} modeled the spiral structure as formed by the overlap of three chunks with different pattern speeds and spatial extent. Specifically, \cite{2dsimulation_spitoni_2} considered three chunks with pattern speeds of 15, 20 and 30 km/s/kpc, respectively. In this context, 10 Galactic rotations would correspond to $\sim$4 Gyr, 3 Gyr and 2 Gyr, respectively. This is coherent with the time needed to explain our observed \cafe\ maps. From collisionless N-body simulations point of view \citep{Saha:2016}, the spiral structures of there simulated disc galaxies have given rise to strong long-lasting two-arm spiral wave modes that persisted for about 5 Gyr with a constant pattern speed.

However, several authors, like \cite{Baba_2009} or \cite{Quillen_2011}, point to the fact that a long co-rotation between the spiral arms and the disc could be very unstable. 
They suggested that spiral arms are not long lived features, but rather, recurrent phenomena. Interestingly, \cite{Grand_2012}, whose simulations are not consistent with long-lived spiral arms, suggest that, in barred spiral galaxies, the central bar might help to maintain the spiral features for longer \citep{Donner_1994, Binney_2008}.

Finally, \cite{Khoperskov_2023}, studied the impact of the local enrichment and pre-existing radial metallicity gradient transformation on the formation of azimuthal metallicity variations in the vicinity of the spiral arms. Their hydrodynamical simulations suggest that, even in the case of co-rotating spirals, the ISM enrichment near the arms alone is not likely responsible for the presence of an azimuthal metallicity pattern, while the key ingredient is a pre-existing radial abundance gradient.


On the other hand, it is interesting to point out that the analysis of the \cafe\ abundance maps for our sample~C, whose age is estimated to be older than 2~Gyr presents also clear \cafe\ deficiencies in the area of the Sag-Car arm and for the Local Arm area at positive azimuths. The \cafe\ features correlate with sample~C metallicity maps from Paper~I, our right panel of Fig. \ref{xymaps_mh_excess} and the azimuthal metallicity variations of RGB stars discussed in \cite{DR3_chemical_cartography}. In addition, the stellar overdensity features, observed by \cite{spiral_arm_radial_action} in a much larger sample of giant stars (dominated by old stars) with \Gaia\ full kinematics, match quite significantly the chemical maps of our sample~C and correlate with some parts of the spiral arms loci of young populations.

To minimize as much as possible the possible presence of young stars in sample C, we performed additional tests, such as modifying the age margin in the Kiel diagram (i.e. shifting the left side of the selected area so that it coincides with older isochrones, i.e. 3, 5 or 10 Gyr), or applying a cut in Galactic azimuthal velocity (i.e. removing stars with $210 <V_{\phi}< 250$ km/s), always obtaining maps consistent with the ones presented in Section \ref{results} (with an even more marked \cafe\ deficiency in the Local arm for negative azimuths.)

Although we expect to have a small fraction of young stars contaminants in our sample C, we cannot completely exclude their presence. We also noted that the $\Teff >$ 4200 K cut removes an important fraction of old (cool) stars. It is therefore possible that the (small) fraction of young contaminants partially contributes to the observed azimuthal variations, making the interpretation of the maps of sample C more challenging.

\begin{figure}[htbp]
\begin{subfigure}{0.5\textwidth}
\includegraphics[width=1\linewidth]{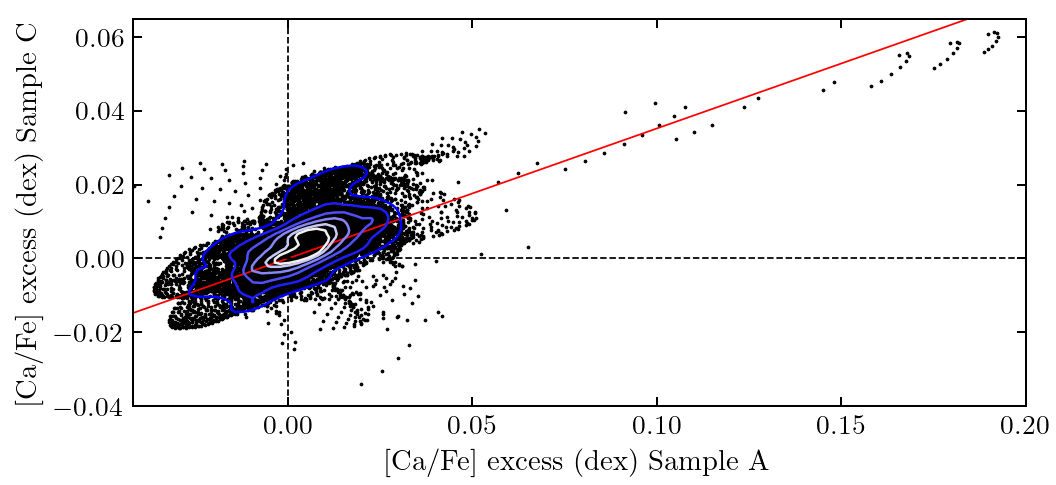} 
\end{subfigure}
\begin{subfigure}{0.5\textwidth}
\includegraphics[width=1\linewidth]{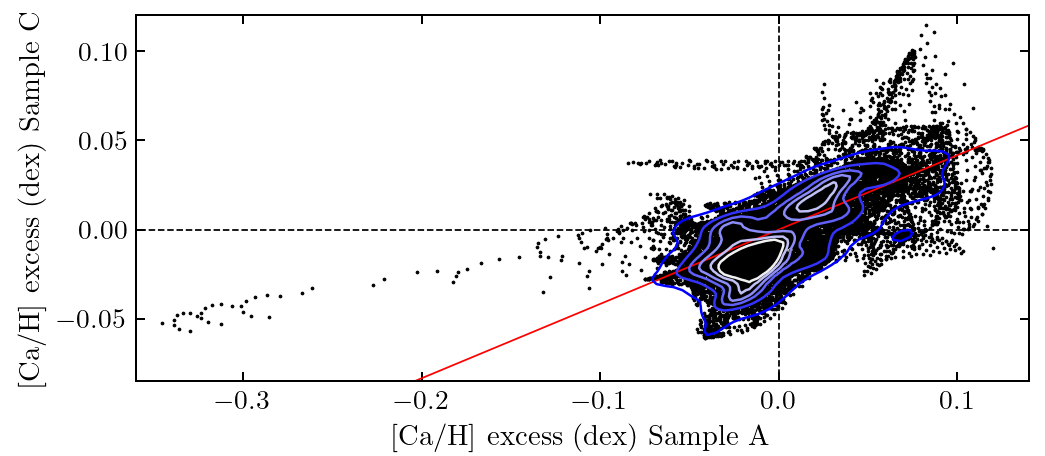} 
\end{subfigure}
    \caption{\textit{Upper panel}: Pixel-to-pixel (black dots) diagram showing the \cafe\ excess variation of sample C as a function of sample A one. Blue contours show the distribution of the \cafe\ excess of sample C versus the one of sample A enclosing fractions of 90, 75, 60, 45, 30, and 20\% of the total number of pixels. The red line shows a linear fit to the black dots. 
    \textit{Bottom panel}: Same as upper panel for the \cah\ excess case.
    }
\label{quantification_AvsC}
\end{figure}

Figure \ref{quantification_AvsC} presents a pixel-to-pixel diagram of the \cafe\ (upper panel) and \cah\ (lower panel) excess comparing samples A and C. We observe that regions most deficient or richest in \cafe\ (identically for \cah) in our young sample have the same behaviour in the older sample, characterised by a Spearman coefficient equal to 0.63 (0.67) corresponding to a strong positive correlation.

However, it is important to recall that nor the \cafe\ maps of the young population (sample~A) or those of the old one (sample~C) present a perfect correlation with the spiral arms loci at all azimuths. The observed \cafe\ deficiencies (or \cah\ enhancements) are stronger for the Sag-Car and Local Arm at positive azimuths, in the direction of the Galactic rotation (for the Local, the Perseus Arm at positive azimuths and the Sag-Car in the case of \cah). On one hand, similar azimuthal variations in the star formation rate have been observed by Herschel \citep{Elia_2022}, suggesting a possible link with the chemical maps with a star formation rate higher for positive azimuths. Similarly, \cite{spiral_arms_dynamic} find evidence of a spiral arm in the form of an overdensity in the dynamically measured disc surface density.
Interestingly, our \cafe\ azimuthal variations show close similarities with their residual fluctuations of the vertical gravitational potential tracing the mass overdensity of the spiral arms. Furthermore, very recently \citep{Hackshaw_apogee_excessmaps} explored chemical gradients and azimuthal substructure in the MW disc with giant stars from the Apache Point Observatory Galactic Evolution Experiment (APOGEE) DR17 data, finding azimuthal variations at the $\pm$0.05 dex-level across the Galactic disc. On the other hand, the observed chemical signatures correlate significantly with several known dust structures, as the Vela Molecular Ridge \citep{extinction_map, vela_cloud} at approximately (X, Y) = (0, -1) kpc (see also Paper~I) than with the spiral arms overdensity contours. Indeed, the dust distribution in the disc is not homogeneous along the spiral arms structure \citep{Lallement22, Vergely22} and it plays a role on the chemical evolution through the so called ''dust cycle'' \citep{Emanuele2017}. A detailed study of the correlation between the \mh\ and \cafe\ maps and the corresponding dust distribution will be presented in a future work (Barbillon et al., in prep.).

It is also important to note that spiral arms are not the only potential drivers of chemical azimuthal variations in Galactic discs. For instance, the bar \citep{DiMatteo:2013,Filion_2023} and radial migration induced by satellites \citep{Carr_22} can also play an important role. This might be particularly important for older stars, which are dynamically relaxed, and had time to experience stellar radial migration. On the other hand, young stars are expected to reflect the motion of the gas from which they were recently born.

Finally, the orbital eccentricities of sample A stars have been considered. Although the stars in this young population are globally on quite circular orbits, the eccentricity distribution obtained through the disc showed clear azimuthal variations
, with a noticeable correlation with the \cafe\ distribution in the Local Arm. Nevertheless, the obtained features are dependent on the considered gravitational potential and a more in-depth study needs to be performed with a larger sample of young stars, less affected by kinematical biases \citep{Poggio:2024}. Furthermore, future studies will consider idealised galaxies simulations \citep{nexus_simulation} to explore in depth the complex physical processes governing the current evolution of our Galaxy. The aim will then be to combine our observational data and simulations to better understand and constrain the structure and evolution of the Galactic disc.

In conclusion, our analysis presents the detection of chemical azimuthal variations in the Galactic disc using $\alpha$-elements in \Gaia\ \gspspec\ data, and reveals a statistically significant correlation between observed chemical inhomogeneities and the position of the Sagittarius-Carina arm, as well as the Local arm in the I and II Galactic quadrants. Our results open new lines of research into the evolutionary processes of the Galactic disc and its interaction with the spiral arms.

\begin{acknowledgements}
    MB thanks the anonymous referee for the useful comments that helped improve the quality of the manuscript. This work presents results from the European Space Agency (ESA) space mission \Gaia\ (\url{https://www.cosmos.esa.int/gaia}). \Gaia\ data are processed by the Gaia Data Processing and Analysis Consortium (DPAC). Funding for the DPAC is provided by national institutions, in particular the institutions participating in the \Gaia\ MultiLateral Agreement (MLA). The \Gaia\ archive website is \url{(https://archives.esac.esa.int/gaia)}. MB and PAP acknowledge financial supports from the French Space Agency, Centre National d’Études Spatiales (CNES). ARB, PdL and ES acknowledge funding from the European Union’s Horizon 2020 research and innovation program under SPACE-H2020 grant agreement number 101004214 (EXPLORE project). This project has received funding from the European Union’s Horizon 2020 research and innovation programme under the Marie Sklodowska-Curie grant agreement N.101063193. We would like to thank the Conseil régional Provence-Alpes-Côte d'Azur for its financial support.
\end{acknowledgements}

\bibliographystyle{aa}  
\bibliography{main} 

\begin{appendix}

\section{Selection criteria}
\label{Appendix_dataselection}

\begin{table*}[]
\centering
\caption{Statistics of sample A and sample C, including for each element the final number of stars, the limiting $\Teff$, the mean signal-to-noise, the abundances dispersion $\sigma$, the median uncertainties, their first and third quartiles distributions.}
\begin{tabular}{cccccccc}
\hline \hline 
\multicolumn{1}{c}{\begin{tabular}[c]{@{}c@{}}Chemical\\ abundance\end{tabular}} & \begin{tabular}[c]{@{}c@{}}Final number \\ of stars\end{tabular} & min $\Teff$ {(}K{)} & Mean SNR & $\sigma$~{(}dex{)} & \begin{tabular}[c]{@{}c@{}}Median \\ uncertainty {(}dex{)}\end{tabular} & \begin{tabular}[c]{@{}c@{}}1st quartile \\  {(}dex{)}\end{tabular}& \begin{tabular}[c]{@{}c@{}}3rd quartile \\ {(}dex{)}\end{tabular} \\ \hline
\hline \\
\multicolumn{8}{c}{\large{Sample A}} \\ \hline
\multicolumn{1}{c|}{\mh} & 13695 & 4200 & 105 & 0.250  & 0.025 & -0.330 & -0.140 \\
\multicolumn{1}{c|}{\cafe} & 11678 & 4200 & 117 & 0.088 & 0.025 & -0.135 & -0.069 \\
\multicolumn{1}{c|}{\cah} & 11678 & 4200 & 117 & 0.191 & 0.050 & -0.414 & -0.250 \\
\multicolumn{1}{c|}{\mgfe} & 689 & 4378 & 515 & 0.087 & 0.045 & -0.131 & -0.044 \\
\hline \\
\multicolumn{8}{c}{\large{Sample C}} \\ \hline
\multicolumn{1}{c|}{\mh}                                                  & 83498                                                            & 4200                     & 96  & 0.182                           & 0.030                                                                   &   -0.440 & -0.190                                                                        \\
\multicolumn{1}{c|}{\cafe}                                          & 74740                                                            & 4200                     & 104  & 0.058                           & 0.025                                                                   & -0.047    &    0.020 \\  
\multicolumn{1}{c|}{\cah} & 74740 & 4200 & 117 & 0.145 & 0.050 & -0.428 & -0.232 \\
\hline
\end{tabular}
\tablefoot{In the case of the \mgfe\, we consider a minimum $\Teff$=4378 K due to the calibration.}
\label{data_selection_details_overfe}
\end{table*}

To obtain the initial \Gaia\ data set for the present analysis, we optimised the procedure described in Appendix~A of Paper I to select precise \gspspec\ $\alpha$-element abundance estimates. To this purpose, we imposed a limit of 0.1~dex in the \alphafe\ abundance uncertainty and we reduced from 0.50~dex to 0.25~dex the maximum allowed uncertainty in \mh. Moreover, the two individual abundance flags (\verb|CaUpLim| and \verb|CaUncer| for \cafe\ and \verb|MgUpLim| and \verb|MgUncer| for \mgfe\ from Table~2 in \cite{DR3_RVS}) are requested to be lower than 2. As a consequence, the definition of the initial working sample is as follows :

\begin{small}
\begin{verbatim}
[M/H]_unc < 0.25 & [$\alpha$/Fe]_unc < 0.1 & vbroadM < 2 
& vradM < 2 & fluxNoise < 4 & extrapol < 3 
& XUpLim < 2 & XUncer < 2 & KMtypestars < 1 
& (astrometric_params_solved = 31 
OR astrometric_params_solved = 95) & ruwe < 1.4 
& duplicated_source = false 
\end{verbatim}
\end{small}

To obtain the \Gaia\ chemo-physical parameters and the kinematics of the selected stars from the \Gaia\ archive, the following query can be submitted (the orange and light blue texts correspond to the calcium and the magnesium specific queries, respectively, and have to be added separately only for studying either the Ca or the Mg):

\begin{scriptsize}
\begin{Verbatim}[commandchars=\\\{\}]

SELECT g.*, ap.teff_gspspec, ap.teff_gspspec_upper, 
ap.teff_gspspec_lower, ap.logg_gspspec, ap.logg_gspspec_upper, 
ap.logg_gspspec_lower, ap.mh_gspspec, ap.mh_gspspec_upper, 
ap.mh_gspspec_lower, \textcolor{orange}{ap.cafe_gspspec, ap.cafe_gspspec_upper}
\textcolor{orange}{ap.cafe_gspspec_lower}, \textcolor{teal}{ap.mgfe_gspspec, ap.mgfe_gspspec_upper}
\textcolor{teal}{ap.mgfe_gspspec_lower}, d.r_med_geo, d.r_lo_geo, 
d.r_hi_geo, c.vz_med, c.vz_lo, c.vz_hi, c.vphi_med, c.vphi_lo, 
c.vphi_hi, c.vrplane_med, c.vrplane_lo, c.vrplane_hi 
FROM gaiadr3.gaia_source AS g INNER JOIN 
gaiadr3.astrophysical_parameters AS ap ON g.source_id = ap.source_id 
INNER JOIN external.gaiaedr3_distance AS d ON
g.source_id = d.source_id INNER JOIN gaiadr3.chemical_cartography 
AS c ON g.source_id = c.source_id
WHERE (((mh_gspspec_upper-mh_gspspec_lower)*0.5 < 0.25) 
AND ((alphafe_gspspec_upper-alphafe_gspspec_lower)*0.5 < 0.1) 
AND ((flags_gspspec LIKE '__0%') OR (flags_gspspec LIKE '__1%')) 
AND ((flags_gspspec LIKE '_____0%') OR (flags_gspspec LIKE '_____1%')) 
AND ((flags_gspspec LIKE '______0%') OR (flags_gspspec LIKE '______1%')
OR (flags_gspspec LIKE '______2%') OR (flags_gspspec LIKE '______3%')) 
AND ((flags_gspspec LIKE '_______0%') 
OR (flags_gspspec LIKE '_______1%') 
OR (flags_gspspec LIKE '_______2%')) 
AND ((flags_gspspec LIKE '____________0%') 
OR (flags_gspspec LIKE '____________0%')) 
\textcolor{orange}{AND ((flags_gspspec LIKE '_____________________0%')}
\textcolor{orange}{OR (flags_gspspec LIKE '_____________________1%')) }
\textcolor{orange}{AND ((flags_gspspec LIKE '______________________0%')}
\textcolor{orange}{OR (flags_gspspec LIKE '______________________1%'))) }
\textcolor{teal}{AND ((flags_gspspec LIKE '_______________0%')} 
\textcolor{teal}{OR (flags_gspspec LIKE '_______________1%')) }
\textcolor{teal}{AND ((flags_gspspec LIKE '________________0%')}
\textcolor{teal}{OR (flags_gspspec LIKE '________________1%'))) }
AND (g.astrometric_params_solved = 31 
OR g.astrometric_params_solved = 95) 
AND g.ruwe < 1.4 AND g.duplicated_source = 'False'

\end{Verbatim}
\end{scriptsize}


The RVS spectra analysed by \gspspec~during DR3 operations were selected to have a SNR > 20 before resampling \citep{DR3_RVS}. In this study, we imposed to select stars having only a SNR > 30 using \verb|rv_expected_sig_to_noise| from the \textbf{gaiadr3.astrophysical$\_$parameters} table.

For the case of the \mgfe\ abundances derived by \gspspec\ from a weak Mg spectral line, we have imposed a threshold in the RVS SNR, selecting stars with \verb|rv_expected_sig_to_noise|$~>250$. This avoids biases in the abundance distribution towards the higher, easier to measure abundance values \citep[see][for a similar case concerning the \gspspec\ \cefe\ abundance]{Contursi23}.
This SNR limit allows to have a median uncertainty in \mgfe\ that is almost twice smaller than the dispersion in the observed \mgfe\ abundances distribution. The final characteristics of the selected stars in samples A and C, taking into account of all the above described criteria, are summarised in Table~\ref{data_selection_details_overfe}.

\begin{figure}[htbp]
\includegraphics[width=1\linewidth]{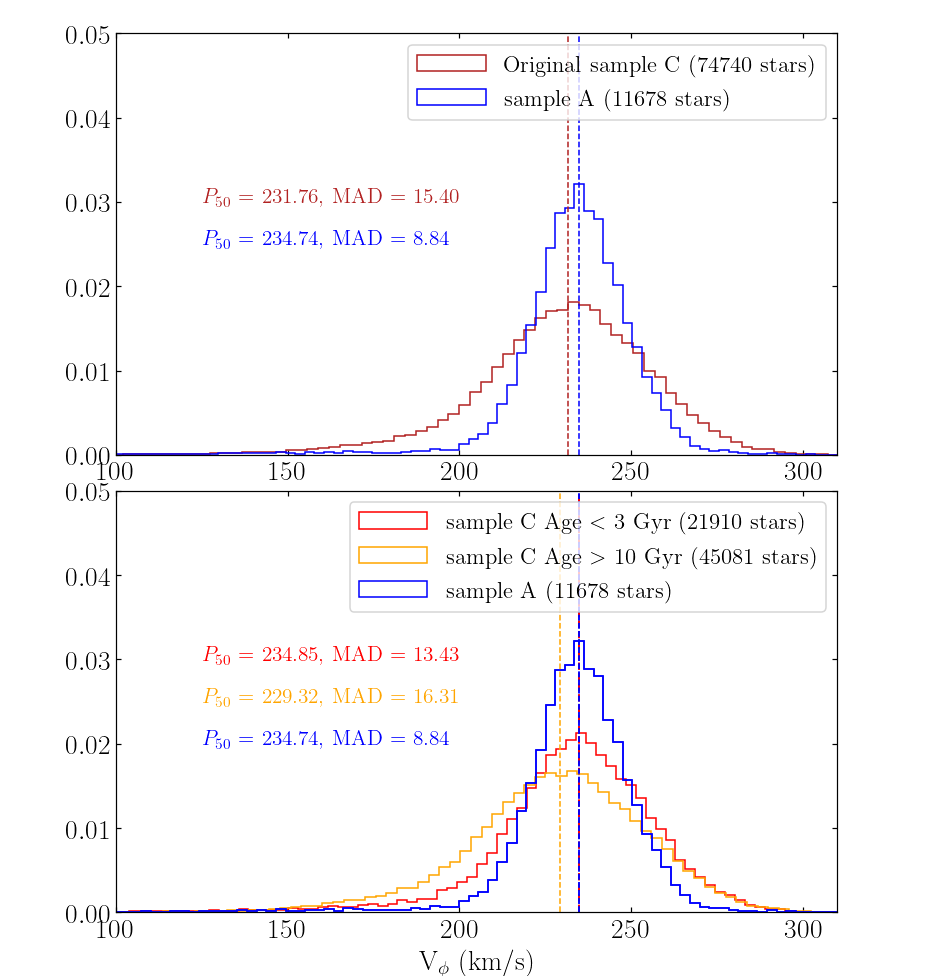}
\caption{\textit{Upper panel:} Probability distribution function of Galactic azimuthal velocities V$_\phi$ for sample A and original sample C of this study, normalised to set the area equal to 1. The dark red bars show the stars distribution of sample C, while blue bars show sample A, displayed with median values (P$_{50}$) and median absolute deviations (MAD) in km/s. Vertical dashed lines denote the median values for each sample. \textit{Bottom panel:} Same as upper panel considering subsets of sample C and sample A. The red and orange bars show the stars distribution of sample C for sources below 3 Gyr and over 10 Gyr, respectively, to avoid contribution from stars in each subset.}
\label{hist_vphi}
\end{figure}

\begin{figure*}[htbp]
\begin{subfigure}{0.5\textwidth}
\includegraphics[width=1\linewidth]{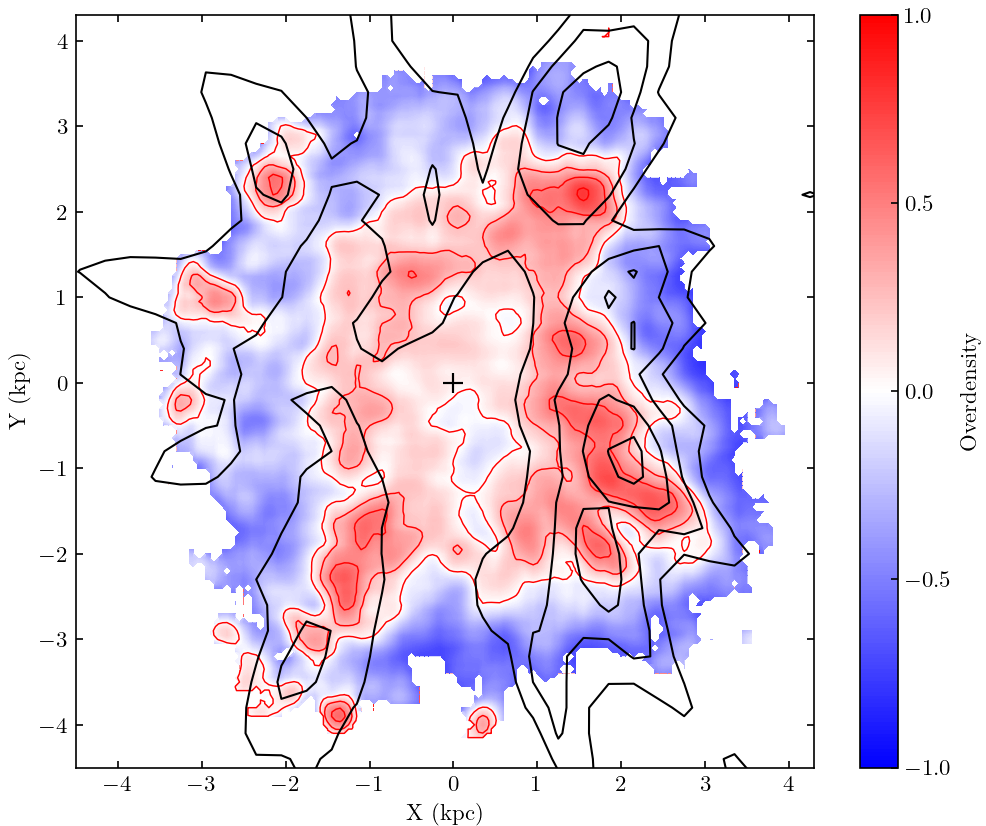}
\end{subfigure}
\begin{subfigure}{0.5\textwidth}
\includegraphics[width=1\linewidth]
{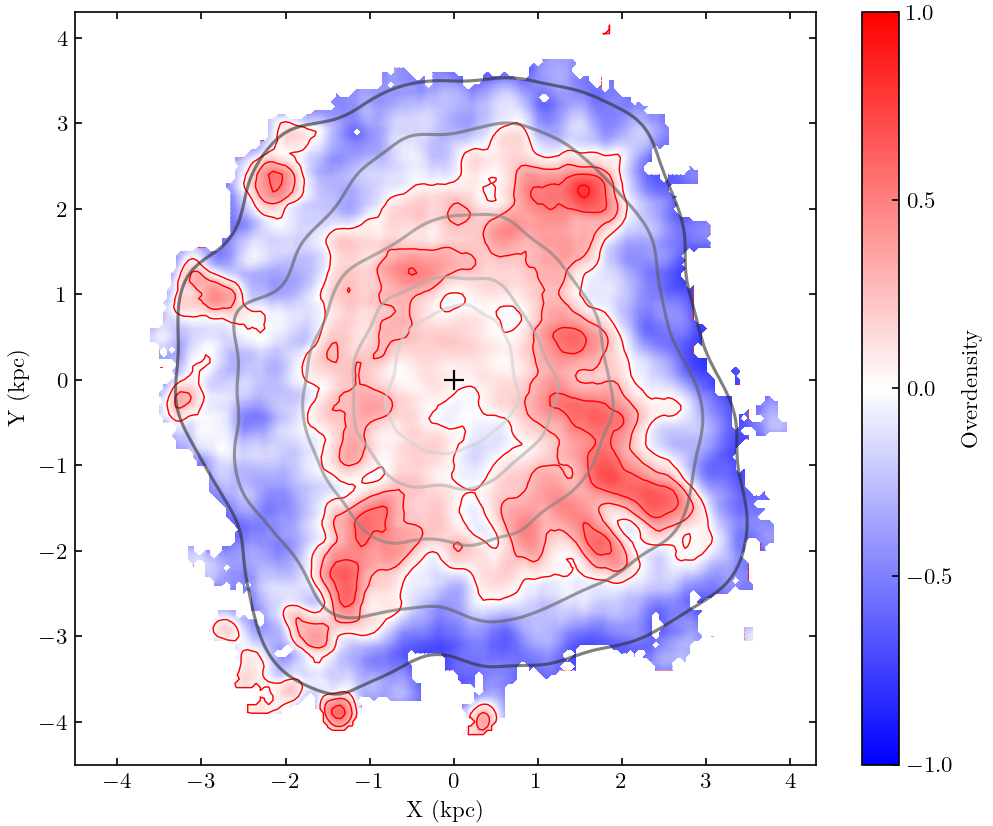}
\end{subfigure}
\caption{ \textit{Left panel:} Overdensity map of sample C with V$_{\phi} \leq$ 210 km/s. Here, 13566 stars are considered. The black contours indicate the position of the spiral arms obtained using the sample of giant stars \citep{spiral_arm_radial_action}. The position of the Sun is indicated by the black cross at (X, Y) = (0, 0) kpc. \textit{Right panel:} Identical to the left panel, with grey circular outlines including stars with $G \leq$ 9, 10, 11, 12 and 13 mag (for the tightest to widest circles).}
\label{xymaps_vphicut_sampleC}
\end{figure*}

Finally, we have explored the spatial distribution of the selected old stars in sample C. To further avoid potential contamination from young stars due to uncertainties in the parameters, we have selected a subsample with V$_{\phi} \leq$ 210 km/s. 
We created the overdensity contours displayed in Fig. \ref{xymaps_vphicut_sampleC}. 
To map these overdensities, we used the procedure described in \cite{spiral_arm_EDR3}, selected a local $h$=300 pc and larger one $h_{larger}=1.8\times h_{local}$ pc. It is important to highlight the fact that this analysis has the advantage of selecting a quite pure old population, using chemo-physical parameters and kinematical filtering, however, the drawback is the lower completeness of the sample. Also, as studied by \cite{CantatGaudin:2024} (cf. Fig. 9, lower left panel) for a fainter red clump sample, the \gspspec\ completeness diminishes as the distance from the Sun increases, mainly as a function of G magnitude (see the right panel of Fig. \ref{xymaps_vphicut_sampleC} for the G magnitude distribution of our sample) and interstellar absorption. Moreover, the completeness might change significantly for different line-of-sights, presumably leading to artifacts when mapping the density distribution of the sources in the Galactic plane. A robust study of the density distribution, and therefore, an evaluation of the sample completeness is out of the scope of this paper. Nevertheless, it is interesting to point out that the obtained density maps presented in the left panel of Fig. \ref{xymaps_vphicut_sampleC} reveals a reasonable match with the \cite{spiral_arm_radial_action} contours. In addition, they show a potential sign of the Perseus Arm in the old population, in agreement with \cite{Khanna2024}, who use a very different procedure and a much more complete sample. Removing the $\Teff$ cut-off in order to increase the size sample (despite the fact that the scanning law is affecting particularly a line-of-sight of the studied sample), it is worth noting that the Perseus signature is always visible and a clearer separation between the Local and Sag-Car arms is discernible.

\end{appendix}

\end{document}